# On the vierbein formalism of general relativity

T. Mei

(Department of Journal, Central China Normal University, Wuhan, Hubei PRO, People's Republic of China

E-Mail:  meitao@mail.ccnu.edu.cn    meitaowh@public.wh.hb.cn )

**Abstract:**   Both the Einstein-Hilbert action and the Einstein equations are discussed under the absolute vierbein formalism. Taking advantage of this form, we prove that the "kinetic energy" term, i.e., the quadratic term of time derivative term, in the Lagrangian of the Einstein-Hilbert action is non-positive definitive. And then, we present two groups of coordinate conditions that lead to positive definitive kinetic energy term in the Lagrangian, as well as the corresponding actions with positive definitive kinetic energy term, respectively. Based on the ADM decomposition, the Hamiltonian representation and canonical quantization of general relativity taking advantage of the actions with positive definitive kinetic energy term are discussed; especially, the Hamiltonian constraints with positive definitive kinetic energy term are given, respectively. Finally, we present a group of gauge conditions such that there is not any second time derivative term in the ten Einstein equations.

There are a number of literatures on the vierbein formalism of general relativity and canonical gravity (See, for example, Refs. [1 ~ 4], and recently, [5, 6]). In this paper, at first, we use absolutely the tetrad field $e_\mu^{\hat\alpha}$ to express the Einstein-Hilbert action and the Einstein equations; and then, discuss some characteristics of the theory. Concretely, taking advantage of this form, we first prove that the "kinetic energy" term, i.e., the quadratic term of time derivative term, in the Lagrangian of the Einstein-Hilbert action is non-positive definitive. And then, we present two groups of coordinate conditions that lead to positive definitive kinetic energy term in the Lagrangian, as well as the corresponding actions with positive definitive kinetic energy term, respectively. Based on the ADM decomposition, the Hamiltonian representation and canonical quantization of general relativity taking advantage of the actions with positive definitive kinetic energy term are discussed; especially, the Hamiltonian constraints with positive definitive kinetic energy term are given, respectively. Finally, we present a group of gauge conditions such that there is not any second time derivative term in the ten Einstein equations.

## 1   The vierbein formalism of general relativity
### 1.1   The vierbein formalism of the Einstein-Hilbert action

As is well known, in general relativity, the Einstein-Hilbert action reads

$$S_{\text{EH}} = \frac{c^3}{16\pi G}\int \sqrt{-g}\,\mathrm{d}^4 x R = \frac{c^3}{16\pi G}\int \sqrt{-g}\,\mathrm{d}^4 x\, L_{\text{g}} + \frac{c^3}{16\pi G}\int \mathrm{d}^4 x\, \frac{\partial}{\partial x^\mu}\left[\sqrt{-g}(g^{\rho\sigma}\Gamma^\mu_{\rho\sigma} - g^{\mu\rho}\Gamma^\sigma_{\rho\sigma})\right], \quad (1\text{-}1)$$

$$L_g = g^{\alpha\beta}(\Gamma^\rho_{\alpha\sigma}\Gamma^\sigma_{\beta\rho} - \Gamma^\rho_{\alpha\beta}\Gamma^\sigma_{\rho\sigma}). \tag{1-2}$$

Although maybe a surface term such as $S_{\text{sur}} = \int_{\partial Y} B\sqrt{h}\,d^3x$ should be added in (1-1)[7], it will be ignored in this paper.

On the other hand, one must use the tetrad field $e^{\hat{\alpha}}_\mu(x)$ to express the Dirac equation in curved spacetime:

$$\left(i\gamma^{\hat{\alpha}} e^\mu_{\hat{\alpha}}(x)\frac{\partial}{\partial x^\mu} + \frac{1}{4}\gamma^{\hat{\gamma}} r_{\hat{\alpha}\hat{\beta}\hat{\gamma}}\sigma^{\hat{\alpha}\hat{\beta}} - \frac{mc}{\hbar}\right)\Psi(x) = 0, \tag{1-3}$$

where the constant matrices $\gamma^{\hat{\alpha}}$ satisfy $\{\gamma^{\hat{\alpha}},\gamma^{\hat{\beta}}\} = -2\eta^{\hat{\alpha}\hat{\beta}}$; $\sigma^{\hat{\alpha}\hat{\beta}} = \frac{i}{2}[\gamma^{\hat{\alpha}},\gamma^{\hat{\beta}}]$; the Ricci's coefficients of rotation

$$r_{\hat{\alpha}\hat{\beta}\hat{\gamma}} = e_{\hat{\alpha}\mu;\nu}e^\mu_{\hat{\beta}} e^\nu_{\hat{\gamma}} = \frac{1}{2}(e_{\hat{\alpha}\mu,\nu} - e_{\hat{\alpha}\nu,\mu})e^\mu_{\hat{\beta}} e^\nu_{\hat{\gamma}} - \frac{1}{2}(e_{\hat{\beta}\mu,\nu} - e_{\hat{\beta}\nu,\mu})e^\mu_{\hat{\alpha}} e^\nu_{\hat{\gamma}} - \frac{1}{2}(e_{\hat{\gamma}\mu,\nu} - e_{\hat{\gamma}\nu,\mu})e^\mu_{\hat{\alpha}} e^\nu_{\hat{\beta}}. \tag{1-4}$$

The relations between metric tensor $g_{\mu\nu}$ and tetrad field $e^{\hat{\alpha}}_\mu$ is as follows

$$g_{\mu\nu} = \eta_{\hat{\alpha}\hat{\beta}} e^{\hat{\alpha}}_\mu e^{\hat{\beta}}_\nu = e^{\hat{\alpha}}_\mu e_{\hat{\alpha}\nu}. \tag{1-5}$$

In this paper, we use the Greek and Latin alphabet without the symbol "^" to denote spacetime indices, and the Greek and Latin alphabet with the symbol "^" to denote local frame components; the Greek alphabet without the symbol "^" is raised or lowered by $g^{\mu\nu}$ or $g_{\mu\nu}$, respectively; the Greek alphabet with the symbol"^" is raised or lowered by $\eta^{\hat{\alpha}\hat{\beta}} = \text{diag}(-1,+1,+1,+1)$ or $\eta_{\hat{\alpha}\hat{\beta}}$, respectively. The value of the Greek and Latin alphabet are 0, 1, 2, 3 and 1, 2, 3, respectively, unless some special statement.

For the purpose that use spacetime indices to express the Dirac equation, we define

$$F_{\mu\nu\lambda} = r_{\hat{\alpha}\hat{\beta}\hat{\gamma}} e^{\hat{\alpha}}_\mu e^{\hat{\beta}}_\nu e^{\hat{\gamma}}_\lambda = e^{\hat{\alpha}}_\mu e_{\hat{\alpha}\nu;\lambda}$$
$$= \frac{1}{2}e^{\hat{\alpha}}_\mu(e_{\hat{\alpha}\nu,\lambda} - e_{\hat{\alpha}\lambda,\nu}) - \frac{1}{2}e^{\hat{\alpha}}_\nu(e_{\hat{\alpha}\mu,\lambda} - e_{\hat{\alpha}\lambda,\mu}) - \frac{1}{2}e^{\hat{\alpha}}_\lambda(e_{\hat{\alpha}\mu,\nu} - e_{\hat{\alpha}\nu,\mu}). \tag{1-6}$$

If $e^{\hat{\alpha}}_\mu$ is changed according to $e^{\hat{\alpha}}_\mu = \frac{\partial \tilde{x}^\nu}{\partial x^\mu}\tilde{e}^{\hat{\alpha}}_\nu$ under coordinate transformation $x^\mu = x^\mu(\tilde{x}^\nu)$, then we can prove that the manner of transformation of $F_{\mu\nu\lambda}$ reads:

$$F_{\mu\nu\lambda} = \frac{\partial \tilde{x}^\alpha}{\partial x^\mu}\frac{\partial \tilde{x}^\beta}{\partial x^\nu}\frac{\partial \tilde{x}^\gamma}{\partial x^\lambda} \tilde{F}_{\alpha\beta\gamma},$$

this means that $F_{\mu\nu\lambda}$ is a three-index covariant tensor in global manifold coordinate system, the index of $F_{\mu\nu\lambda}$ is thus raised or lowered by $g^{\mu\nu}$ or $g_{\mu\nu}$, respectively. For example,

$$F^{\rho\sigma}{}_\sigma = g^{\rho\mu}g^{\nu\lambda}F_{\mu\nu\lambda} = g^{\rho\nu}e^\mu_{\hat{\alpha}}(e^{\hat{\alpha}}_{\mu,\nu} - e^{\hat{\alpha}}_{\nu,\mu}). \tag{1-7}$$

Similar to the characteristic that $r_{\hat{\alpha}\hat{\beta}\hat{\gamma}}$ is antisymmetric in the first pair of indices:

$$r_{\hat{\alpha}\hat{\beta}\hat{\gamma}} = -r_{\hat{\beta}\hat{\alpha}\hat{\gamma}}, \qquad (1\text{-}8)$$

$F_{\mu\nu\lambda}$ is also antisymmetric in the first pair of indices:

$$F_{\mu\nu\lambda} = -F_{\nu\mu\lambda}. \qquad (1\text{-}9)$$

Hence, both $r_{\hat{\alpha}\hat{\beta}\hat{\gamma}}$ and $F_{\mu\nu\lambda}$ have at most 24 independent components in 4-dimensional Riemannian geometry, respectively.

Using $F_{\mu\nu\lambda}$, the Dirac equation (1-3) and the corresponding action can be expressed as follows:

$$\left( i\gamma^\mu(x)\frac{\partial}{\partial x^\mu} + \frac{1}{4}\gamma^\mu(x) F_{\alpha\beta\mu}\sigma^{\alpha\beta}(x) - \frac{mc}{\hbar} \right)\Psi(x) = 0, \qquad (1\text{-}10)$$

$$\begin{aligned}
S_{\mathrm{D}} &= \int \sqrt{-g}\, \mathrm{d}^4 x L_{\mathrm{D}} = \int \left| {}^4 e \right| \mathrm{d}^4 x\, L_{\mathrm{D}}, \\
L_{\mathrm{D}} &= \overline{\Psi}(x)\left( i\hbar\gamma^\mu(x)\frac{\partial}{\partial x^\mu} + \frac{1}{4}\hbar\gamma^\mu(x) F_{\alpha\beta\mu}\sigma^{\alpha\beta}(x) - mc \right)\Psi(x),
\end{aligned} \qquad (1\text{-}11)$$

where $\gamma^\mu(x) = \gamma^{\hat{\alpha}} e^\mu_{\hat{\alpha}}$, $\left| {}^4 e \right| = \det\left[ e^{\hat{\alpha}}_\mu \right]$ is the determinant of the $4\times 4$ matrix $\left[ e^{\hat{\alpha}}_\mu \right]$.

Now that the Dirac equation in curved spacetime must be expressed by tetrad field $e^{\hat{\alpha}}_\mu$, we should find out a Lagrangian on tetrad field $e^{\hat{\alpha}}_\mu$ or $F_{\mu\nu\lambda}$. If we use the method of the Yang-Mills-like field, then we can define a "covariant derivative" $D_\mu = \partial_\mu - \frac{i}{4} F_{\hat{\alpha}\hat{\beta}\mu}\sigma^{\hat{\alpha}\hat{\beta}}$, where $F_{\hat{\alpha}\hat{\beta}\mu} = e^\rho_{\hat{\alpha}} e^\sigma_{\hat{\beta}} F_{\rho\sigma\mu}$, the corresponding field strength reads

$$F_{(1)\mu\nu} = i[D_\mu, D_\nu] = \frac{1}{4}\sigma^{\hat{\alpha}\hat{\beta}}\left( \partial_\mu F_{\hat{\alpha}\hat{\beta}\nu} - \partial_\nu F_{\hat{\alpha}\hat{\beta}\mu} \right) + \frac{1}{2}\sigma^{\hat{\alpha}\hat{\beta}}\eta^{\hat{\rho}\hat{\sigma}} F_{\hat{\rho}\hat{\alpha}\mu} F_{\hat{\sigma}\hat{\beta}\nu},$$

and we can prove that the corresponding Lagrangian

$$L_{(1)} = -\frac{1}{4}\mathrm{Tr}\!\left( F^{\mu\nu}_{(1)} F_{(1)\mu\nu} \right) = -\frac{1}{8} R^{\mu\nu\rho\sigma} R_{\mu\nu\rho\sigma},$$

where $R_{\mu\nu\rho\sigma}$ is the Riemann curvature tensor (The relation between $R_{\mu\nu\rho\sigma}$ and $F_{\mu\nu\lambda}$ will be given in below (1-14)). We see that this theory includes higher-derivative and is not equivalent to general relativity[8].

Y.-S. Duan and J.-Y. Zhang[9] have given a Lagrangian that can leads to the Einstein equations:

$$L_{\mathrm{G}} = r^{\hat{\alpha}\hat{\beta}}{}_{\hat{\beta}} r_{\hat{\alpha}}{}^{\hat{\gamma}}{}_{\hat{\gamma}} - r^{\hat{\alpha}\hat{\beta}\hat{\gamma}} r_{\hat{\alpha}\hat{\gamma}\hat{\beta}}; \qquad (1\text{-}12)$$

And the remainder of $L_g$ and $L_{\mathrm{G}}$ is a total derivative:

$$L_g = L_{\mathrm{G}} + \frac{\partial}{\partial x^\mu}\left[ \sqrt{-g}\left( e^{\hat{\alpha}\nu}\frac{\partial e^\mu_{\hat{\alpha}}}{\partial x^\nu} - e^{\hat{\alpha}\mu}\frac{\partial e^\nu_{\hat{\alpha}}}{\partial x^\nu} \right) \right]. \qquad (1\text{-}13)$$

In fact, from $e^{\mu}_{\hat{\alpha};\sigma;\nu} - e^{\mu}_{\hat{\alpha};\nu;\sigma} = e^{\lambda}_{\hat{\alpha}} R^{\mu}_{\lambda\nu\sigma}$ we have

$$\begin{aligned}
R^{\mu}_{\rho\nu\sigma} &= \Gamma^{\mu}_{\rho\sigma,\nu} - \Gamma^{\mu}_{\rho\nu,\sigma} + \Gamma^{\mu}_{\lambda\nu}\Gamma^{\lambda}_{\rho\sigma} - \Gamma^{\mu}_{\lambda\sigma}\Gamma^{\lambda}_{\rho\nu} \\
&= e^{\hat{\alpha}}_{\rho}(e^{\mu}_{\hat{\alpha};\sigma;\nu} - e^{\mu}_{\hat{\alpha};\nu;\sigma}) \\
&= \left(e^{\hat{\alpha}}_{\rho} e^{\mu}_{\hat{\alpha};\sigma}\right)_{;\nu} - \left(e^{\hat{\alpha}}_{\rho} e^{\mu}_{\hat{\alpha};\nu}\right)_{;\sigma} - e^{\hat{\alpha}}_{\rho,\nu} e^{\mu}_{\hat{\alpha};\sigma} + e^{\hat{\alpha}}_{\rho,\sigma} e^{\mu}_{\hat{\alpha};\nu} \\
&= F_{\rho}{}^{\mu}{}_{\sigma;\nu} - F_{\rho}{}^{\mu}{}_{\nu;\sigma} + F_{\lambda\rho\sigma} F^{\lambda\mu}{}_{\nu} - F_{\lambda\rho\nu} F^{\lambda\mu}{}_{\sigma},
\end{aligned} \quad (1\text{-}14)$$

and, further,

$$\begin{aligned}
R &= g^{\rho\sigma} R^{\lambda}_{\rho\lambda\sigma} = F^{\rho\sigma}{}_{\rho;\sigma} - F^{\rho\sigma}{}_{\sigma;\rho} + F^{\rho\sigma}{}_{\sigma} F_{\rho\lambda}{}^{\lambda} - F^{\rho\sigma\lambda} F_{\rho\lambda\sigma} \\
&= -2 \frac{1}{\sqrt{-g}} \frac{\partial}{\partial x^{\rho}} (\sqrt{-g} F^{\rho\sigma}{}_{\sigma}) + F^{\rho\sigma}{}_{\sigma} F_{\rho\lambda}{}^{\lambda} - F^{\rho\sigma\lambda} F_{\rho\lambda\sigma}.
\end{aligned} \quad (1\text{-}15)$$

Substituting the above formula to (1-1), we obtain

$$S_{\text{EH}} = \frac{c^3}{16\pi G} \int \sqrt{-g}\, d^4 x R = \frac{c^3}{16\pi G} \int \left|{}^4 e\right| d^4 x L_{\text{G}} + \frac{c^3}{16\pi G} \int d^4 x \frac{\partial}{\partial x^{\rho}} (-2\sqrt{-g} F^{\rho\sigma}{}_{\sigma}), \quad (1\text{-}16)$$

$$\begin{aligned}
L_{\text{G}} &= F^{\rho\sigma}{}_{\sigma} F_{\rho\lambda}{}^{\lambda} - F^{\rho\sigma\lambda} F_{\rho\lambda\sigma} = r^{\hat{\alpha}\hat{\beta}}{}_{\hat{\beta}} r_{\hat{\alpha}}{}^{\hat{\gamma}}{}_{\hat{\gamma}} - r^{\hat{\alpha}\hat{\beta}\hat{\gamma}} r_{\hat{\alpha}\hat{\gamma}\hat{\beta}} \\
&= -\frac{1}{4}\left(\eta_{\hat{\alpha}\hat{\beta}} g^{\mu\rho} g^{\nu\sigma} + 2 g^{\mu\rho} e^{\sigma}_{\hat{\alpha}} e^{\nu}_{\hat{\beta}} - 4 g^{\mu\rho} e^{\nu}_{\hat{\alpha}} e^{\sigma}_{\hat{\beta}}\right)\left(e^{\hat{\alpha}}_{\mu,\nu} - e^{\hat{\alpha}}_{\nu,\mu}\right)\left(e^{\hat{\beta}}_{\rho,\sigma} - e^{\hat{\beta}}_{\sigma,\rho}\right).
\end{aligned} \quad (1\text{-}17)$$

In (1-17), $g^{\mu\nu} = e^{\hat{\alpha}\mu} e^{\nu}_{\hat{\alpha}}$. Notice that there is not any term that is higher than first derivative in $L_{\text{G}}$, and $L_{\text{G}}$ is a scalar under coordinate transformation $x^{\mu} = x^{\mu}(\tilde{x}^{\nu})$.

If we define

$$\begin{aligned}
F_{(2)\lambda\mu\nu} &= \frac{\sqrt{2}}{3}\left[-2 F_{\mu\nu\lambda} + \left(F_{\lambda\mu\nu} - F_{\lambda\nu\mu}\right) - \left(g_{\lambda\mu} F_{\nu\tau}{}^{\tau} - g_{\lambda\nu} F_{\mu\tau}{}^{\tau}\right)\right] \\
&+ i\frac{2}{3}\left[\delta_1 F_{\mu\nu\lambda} + \delta_1\left(F_{\lambda\mu\nu} - F_{\lambda\nu\mu}\right) + \delta_2 \left(g_{\lambda\mu} F_{\nu\tau}{}^{\tau} - g_{\lambda\nu} F_{\mu\tau}{}^{\tau}\right)\right],
\end{aligned}$$

where $\delta_1 = \pm 1$, $\delta_2 = \pm 1$, then we can prove

$$L_{\text{G}} = -\frac{1}{4} F^{\lambda\mu\nu}_{(2)} F_{(2)\lambda\mu\nu} = -\frac{1}{4} F^{\hat{\alpha}\mu\nu}_{(2)} F_{(2)\hat{\alpha}\mu\nu},$$

where $F_{(2)\hat{\alpha}\mu\nu} = e^{\lambda}_{\hat{\alpha}} F_{(2)\lambda\mu\nu}$. Because $F_{(2)\hat{\alpha}\mu\nu} = -F_{(2)\hat{\alpha}\nu\mu}$, the above expression is quite similar to the form of the Lagrangian of the Yang-Mills field; however, it is essentially not equivalent to the Yang-Mills field due to $i = \sqrt{-1}$ appears in the field strength $F_{(2)\lambda\mu\nu}$ inevitably.

## 1.2 The characteristics of transformation of $L_{\text{G}}$ under local Lorentz transformation

Because $F_{\mu\nu\lambda}$ is a three-index covariant tensor for the global manifold coordinate system, and notice that $F_{\mu\nu\lambda}$ satisfies (1-9), the most general form of scalar constructed by the quadratic terms of $F_{\mu\nu\lambda}$ reads:

$$L(F_{\mu\nu\lambda}) = \kappa + aF^{\mu\nu}{}_{\nu} F_{\mu\lambda}{}^{\lambda} + bF^{\mu\nu\lambda} F_{\mu\lambda\nu} + \varepsilon F^{\mu\nu\lambda} F_{\mu\nu\lambda}. \tag{1-18}$$

where $\kappa$ is a constant.

Under local Lorentz transformation, the manner of transformation of $e_\mu^{\hat{\alpha}}$ reads:

$$e_\mu^{\hat{\alpha}} = \Lambda_{\hat{\beta}}^{\hat{\alpha}}(x)\tilde{e}_\mu^{\hat{\beta}}, \quad e_{\hat{\alpha}}^\mu = \overline{\Lambda}_{\hat{\alpha}}^{\hat{\beta}}(x)\tilde{e}_{\hat{\beta}}^\mu, \tag{1-19}$$

where $\Lambda_{\hat{\beta}}^{\hat{\alpha}} \overline{\Lambda}_{\hat{\alpha}}^{\hat{\gamma}} = \delta_{\hat{\beta}}^{\hat{\gamma}}$, $\Lambda_{\hat{\alpha}}^{\hat{\gamma}} \overline{\Lambda}_{\hat{\beta}}^{\hat{\alpha}} = \delta_{\hat{\beta}}^{\hat{\gamma}}$. We can obtain the rule of the transformation of $F_{\mu\nu\lambda}$ according to (1-6) and (1-19):

$$\begin{aligned} F_{\mu\nu\lambda} &= \tilde{F}_{\mu\nu\lambda} + \Lambda_{\hat{\beta}}^{\hat{\alpha}} \overline{\Lambda}_{\hat{\alpha},\lambda}^{\hat{\gamma}} \tilde{e}_\mu^{\hat{\beta}} \tilde{e}_{\hat{\gamma}\nu}, \\ \tilde{F}_{\mu\nu\lambda} &= \tilde{e}_\mu^{\hat{\alpha}} \tilde{e}_{\hat{\alpha}\nu;\lambda}. \end{aligned} \tag{1-20}$$

And we have

$$\int |{}^4 e| \mathrm{d}^4 \boldsymbol{x} L(F_{\mu\nu\lambda}) = \int |{}^4 e| \mathrm{d}^4 \boldsymbol{x} L(\tilde{F}_{\mu\nu\lambda}) + \int \Delta \mathrm{d}^4 \boldsymbol{x},$$

$$\Delta = a\Delta_a + b\Delta_b + \varepsilon\Delta_\varepsilon;$$

$$\Delta_a = \sqrt{-g}(2\Lambda_{\hat{\beta}}^{\hat{\alpha}} \overline{\Lambda}_{\hat{\alpha},\sigma}^{\hat{\gamma}} \tilde{e}_\mu^{\hat{\beta}} \tilde{e}_{\hat{\gamma}}^\sigma \tilde{F}^{\mu\rho}{}_\rho + \Lambda_{\hat{\beta}}^{\hat{\alpha}} \overline{\Lambda}_{\hat{\alpha},\rho}^{\hat{\gamma}} \tilde{e}^{\hat{\beta}\mu} \tilde{e}_{\hat{\gamma}}^\rho \Lambda_{\hat{\chi}}^{\hat{\theta}} \overline{\Lambda}_{\hat{\theta},\sigma}^{\hat{\tau}} \tilde{e}_\mu^{\hat{\chi}} \tilde{e}_{\hat{\tau}}^\sigma),$$

$$\Delta_b = \sqrt{-g}(2\Lambda_{\hat{\beta}}^{\hat{\alpha}} \overline{\Lambda}_{\hat{\alpha},\nu}^{\hat{\gamma}} \tilde{e}_\mu^{\hat{\beta}} \tilde{e}_{\hat{\gamma}\lambda} \tilde{F}^{\mu\nu\lambda} + \Lambda_{\hat{\beta}}^{\hat{\alpha}} \overline{\Lambda}_{\hat{\alpha}}^{\hat{\gamma},\lambda} \tilde{e}^{\hat{\beta}\mu} \tilde{e}_{\hat{\gamma}}^\nu \Lambda_{\hat{\chi}}^{\hat{\theta}} \overline{\Lambda}_{\hat{\theta},\nu}^{\hat{\tau}} \tilde{e}_\mu^{\hat{\chi}} \tilde{e}_{\hat{\tau}\lambda}),$$

$$\Delta_\varepsilon = \sqrt{-g}(2\Lambda_{\hat{\beta}}^{\hat{\alpha}} \overline{\Lambda}_{\hat{\alpha},\lambda}^{\hat{\gamma}} \tilde{e}_\mu^{\hat{\beta}} \tilde{e}_{\hat{\gamma}\nu} \tilde{F}^{\mu\nu\lambda} + \Lambda_{\hat{\beta}}^{\hat{\alpha}} \overline{\Lambda}_{\hat{\alpha}}^{\hat{\gamma},\lambda} \tilde{e}^{\hat{\beta}\mu} \tilde{e}_{\hat{\gamma}}^\nu \Lambda_{\hat{\chi}}^{\hat{\theta}} \overline{\Lambda}_{\hat{\theta},\lambda}^{\hat{\tau}} \tilde{e}_\mu^{\hat{\chi}} \tilde{e}_{\hat{\tau}\nu}).$$

Using the formulas on the Lorentz transformation:

$$\eta^{\hat{\alpha}\hat{\gamma}} \Lambda_{\hat{\gamma}}^{\hat{\beta}} = \eta^{\hat{\beta}\hat{\gamma}} \overline{\Lambda}_{\hat{\gamma}}^{\hat{\alpha}}, \quad \eta_{\hat{\alpha}\hat{\gamma}} \Lambda_{\hat{\beta}}^{\hat{\gamma}} = \eta_{\hat{\beta}\hat{\gamma}} \overline{\Lambda}_{\hat{\alpha}}^{\hat{\gamma}},$$

which can be obtained by the definition of the Lorentz transformation $\eta^{\hat{\rho}\hat{\sigma}} \Lambda_{\hat{\rho}}^{\hat{\alpha}} \Lambda_{\hat{\sigma}}^{\hat{\beta}} = \eta^{\hat{\alpha}\hat{\beta}}$, for $\Delta_a$ we have:

$$\begin{aligned} \Delta_a &= 2\sqrt{-g} \Lambda_{\hat{\beta}}^{\hat{\alpha}} \overline{\Lambda}_{\hat{\alpha},\rho}^{\hat{\gamma}} \tilde{e}_{\hat{\gamma}}^\rho \tilde{e}^{\hat{\beta}\sigma}{}_{,\sigma} + 2\frac{\partial \sqrt{-g}}{\partial x^\sigma} \Lambda_{\hat{\beta}}^{\hat{\alpha}} \overline{\Lambda}_{\hat{\alpha},\rho}^{\hat{\gamma}} \tilde{e}_{\hat{\gamma}}^\rho \tilde{e}^{\hat{\beta}\sigma} + \sqrt{-g}\eta^{\hat{\beta}\hat{\chi}} \Lambda_{\hat{\beta}}^{\hat{\alpha}} \Lambda_{\hat{\chi}}^{\hat{\theta}} \overline{\Lambda}_{\hat{\alpha},\rho}^{\hat{\gamma}} \overline{\Lambda}_{\hat{\theta},\sigma}^{\hat{\tau}} \tilde{e}_{\hat{\gamma}}^\rho \tilde{e}_{\hat{\tau}}^\sigma \\ &= 2\frac{\partial}{\partial x^\sigma}\left(\sqrt{-g} \Lambda_{\hat{\beta}}^{\hat{\alpha}} \overline{\Lambda}_{\hat{\alpha},\rho}^{\hat{\gamma}} \tilde{e}_{\hat{\gamma}}^\rho \tilde{e}^{\hat{\beta}\sigma}\right) - 2\sqrt{-g} \Lambda_{\hat{\beta}}^{\hat{\alpha}} \overline{\Lambda}_{\hat{\alpha},\rho}^{\hat{\gamma}} \tilde{e}_{\hat{\gamma},\sigma}^\rho \tilde{e}^{\hat{\beta}\sigma} + \sqrt{-g} \Lambda_{\hat{\beta},\rho}^{\hat{\alpha}} \overline{\Lambda}_{\hat{\alpha},\sigma}^{\hat{\gamma}} \tilde{e}_{\hat{\gamma}}^\rho \tilde{e}^{\hat{\beta}\sigma} - \Delta_{a1}, \end{aligned}$$

where

$$\begin{aligned} \Delta_{a1} &= 2\sqrt{-g} \Lambda_{\hat{\beta}}^{\hat{\alpha}} \overline{\Lambda}_{\hat{\alpha},\rho,\sigma}^{\hat{\gamma}} \tilde{e}_{\hat{\gamma}}^\rho \tilde{e}^{\hat{\beta}\sigma} + \sqrt{-g} \Lambda_{\hat{\beta},\sigma}^{\hat{\alpha}} \overline{\Lambda}_{\hat{\alpha},\rho}^{\hat{\gamma}} \tilde{e}_{\hat{\gamma}}^\rho \tilde{e}^{\hat{\beta}\sigma} + \sqrt{-g} \Lambda_{\hat{\beta},\rho}^{\hat{\alpha}} \overline{\Lambda}_{\hat{\alpha},\sigma}^{\hat{\gamma}} \tilde{e}_{\hat{\gamma}}^\rho \tilde{e}^{\hat{\beta}\sigma} \\ &= \sqrt{-g}(\Lambda_{\hat{\beta}}^{\hat{\alpha}} \overline{\Lambda}_{\hat{\alpha}}^{\hat{\gamma}})_{,\rho,\sigma} \tilde{e}_{\hat{\gamma}}^\rho \tilde{e}^{\hat{\beta}\sigma} = \sqrt{-g}(\delta_{\hat{\beta}}^{\hat{\gamma}})_{,\rho,\sigma} \tilde{e}_{\hat{\gamma}}^\rho \tilde{e}^{\hat{\beta}\sigma} = 0. \end{aligned}$$

For $\Delta_b$, notice $\Lambda_{\hat{\beta}}^{\hat{\alpha}} \overline{\Lambda}_{\hat{\alpha},\nu}^{\hat{\gamma}} = -\Lambda_{\hat{\beta},\nu}^{\hat{\alpha}} \overline{\Lambda}_{\hat{\alpha}}^{\hat{\gamma}}$ which can be obtained by $(\Lambda_{\hat{\beta}}^{\hat{\alpha}} \overline{\Lambda}_{\hat{\alpha}}^{\hat{\gamma}})_{,\nu} = (\delta_{\hat{\beta}}^{\hat{\gamma}})_{,\nu} = 0$ we have

$$\Delta_b = -2\sqrt{-g} \Lambda_{\hat{\beta}}^{\hat{\alpha}} \overline{\Lambda}_{\hat{\alpha},\rho}^{\hat{\gamma}} \tilde{e}_{\hat{\gamma},\sigma}^\rho \tilde{e}^{\hat{\beta}\sigma} + 2\sqrt{-g} \Lambda_{\hat{\beta}}^{\hat{\alpha}} \overline{\Lambda}_{\hat{\alpha},\rho}^{\hat{\gamma}} \tilde{e}_{\hat{\gamma}}^\sigma \tilde{e}^{\hat{\beta}\lambda} \Gamma_{\lambda\sigma}^\rho + \sqrt{-g} \Lambda_{\hat{\beta},\rho}^{\hat{\alpha}} \overline{\Lambda}_{\hat{\alpha},\sigma}^{\hat{\gamma}} \tilde{e}_{\hat{\gamma}}^\rho \tilde{e}^{\hat{\beta}\sigma},$$

As for the middle term in the above expression, due to

$$\Lambda^{\hat{\alpha}}_{\hat{\beta}}\overline{\Lambda}^{\hat{\gamma}}_{\hat{\alpha},\rho}\widetilde{e}^{\sigma}_{\hat{\gamma}}\widetilde{e}^{\hat{\beta}\lambda}\Gamma^{\rho}_{\lambda\sigma} = \eta^{\hat{\beta}\hat{\chi}}\Lambda^{\hat{\alpha}}_{\hat{\beta}}\eta_{\hat{\gamma}\hat{\tau}}\overline{\Lambda}^{\hat{\gamma}}_{\hat{\alpha},\rho}\widetilde{e}^{\hat{\tau}\sigma}\widetilde{e}^{\lambda}_{\hat{\chi}}\Gamma^{\rho}_{\lambda\sigma} = \eta^{\hat{\alpha}\hat{\beta}}\overline{\Lambda}^{\hat{\chi}}_{\hat{\beta}}\eta_{\hat{\alpha}\hat{\gamma}}\Lambda^{\hat{\gamma}}_{\hat{\tau},\rho}\widetilde{e}^{\hat{\tau}\sigma}\widetilde{e}^{\lambda}_{\hat{\chi}}\Gamma^{\rho}_{\lambda\sigma}$$
$$= \Lambda^{\hat{\alpha}}_{\hat{\beta},\rho}\overline{\Lambda}^{\hat{\gamma}}_{\hat{\alpha}}\widetilde{e}^{\hat{\beta}\sigma}\widetilde{e}^{\lambda}_{\hat{\gamma}}\Gamma^{\rho}_{\lambda\sigma} = \Lambda^{\hat{\alpha}}_{\hat{\beta},\rho}\overline{\Lambda}^{\hat{\gamma}}_{\hat{\alpha}}\widetilde{e}^{\hat{\beta}\lambda}\widetilde{e}^{\sigma}_{\hat{\gamma}}\Gamma^{\rho}_{\sigma\lambda},$$

we have

$$2\sqrt{-g}\Lambda^{\hat{\alpha}}_{\hat{\beta}}\overline{\Lambda}^{\hat{\gamma}}_{\hat{\alpha},\rho}\widetilde{e}^{\sigma}_{\hat{\gamma}}\widetilde{e}^{\hat{\beta}\lambda}\Gamma^{\rho}_{\lambda\sigma} = \sqrt{-g}\Lambda^{\hat{\alpha}}_{\hat{\beta}}\overline{\Lambda}^{\hat{\gamma}}_{\hat{\alpha},\rho}\widetilde{e}^{\sigma}_{\hat{\gamma}}\widetilde{e}^{\hat{\beta}\lambda}\Gamma^{\rho}_{\lambda\sigma} + \sqrt{-g}\Lambda^{\hat{\alpha}}_{\hat{\beta},\rho}\overline{\Lambda}^{\hat{\gamma}}_{\hat{\alpha}}\widetilde{e}^{\hat{\beta}\lambda}\widetilde{e}^{\sigma}_{\hat{\gamma}}\Gamma^{\rho}_{\sigma\lambda}$$
$$= \sqrt{-g}(\Lambda^{\hat{\alpha}}_{\hat{\beta}}\overline{\Lambda}^{\hat{\gamma}}_{\hat{\alpha},\rho} + \Lambda^{\hat{\alpha}}_{\hat{\beta},\rho}\overline{\Lambda}^{\hat{\gamma}}_{\hat{\alpha}})\widetilde{e}^{\sigma}_{\hat{\gamma}}\widetilde{e}^{\hat{\beta}\lambda}\Gamma^{\rho}_{\lambda\sigma} = \sqrt{-g}(\Lambda^{\hat{\alpha}}_{\hat{\beta}}\overline{\Lambda}^{\hat{\gamma}}_{\hat{\alpha}})_{,\rho}\widetilde{e}^{\sigma}_{\hat{\gamma}}\widetilde{e}^{\hat{\beta}\lambda}\Gamma^{\rho}_{\lambda\sigma}$$
$$= \sqrt{-g}(\delta^{\hat{\gamma}}_{\hat{\beta}})_{,\rho}\widetilde{e}^{\sigma}_{\hat{\gamma}}\widetilde{e}^{\hat{\beta}\lambda}\Gamma^{\rho}_{\lambda\sigma} = 0.$$

For $\Delta_\varepsilon$ we have

$$\Delta_\varepsilon = 2\sqrt{-g}\Lambda^{\hat{\alpha}}_{\hat{\beta}}\overline{\Lambda}^{\hat{\gamma},\rho}_{\hat{\alpha}}\widetilde{e}^{\hat{\beta}\sigma}_{,\rho}\widetilde{e}_{\hat{\gamma}\sigma} + 2\sqrt{-g}\Lambda^{\hat{\alpha}}_{\hat{\beta}}\overline{\Lambda}^{\hat{\gamma},\rho}_{\hat{\alpha}}\widetilde{e}^{\hat{\beta}\lambda}\widetilde{e}_{\hat{\gamma}\sigma}\Gamma^{\sigma}_{\lambda\rho} + \sqrt{-g}\Lambda^{\hat{\alpha}}_{\hat{\beta},\lambda}\overline{\Lambda}^{\hat{\beta},\lambda}_{\hat{\alpha}}$$
$$= \sqrt{-g}\left\{ 2\left[\Lambda^{\hat{\alpha}}_{\hat{\beta}}\overline{\Lambda}^{\hat{\gamma}}_{\hat{\alpha},\rho}\widetilde{e}^{\hat{\beta}\rho}_{,\sigma}\widetilde{e}^{\sigma}_{\hat{\gamma}} - \Lambda^{\hat{\alpha}}_{\hat{\beta}}\overline{\Lambda}^{\hat{\gamma},\rho}_{\hat{\alpha}}(\widetilde{e}^{\hat{\beta}}_{\sigma}\widetilde{e}^{\sigma}_{\hat{\gamma},\rho} + \widetilde{e}^{\sigma}_{\hat{\gamma}}\widetilde{e}^{\hat{\beta}}_{\rho,\sigma})\right] + \Lambda^{\hat{\alpha}}_{\hat{\beta},\lambda}\overline{\Lambda}^{\hat{\beta},\lambda}_{\hat{\alpha}}\right\}.$$

Summarizing the above results, we obtain

$$\int \Delta d^4 x = 2a\int \frac{\partial}{\partial x^\mu}\left(\sqrt{-g}\Lambda^{\hat{\alpha}}_{\hat{\beta}}\overline{\Lambda}^{\hat{\gamma}}_{\hat{\alpha},\lambda}\widetilde{e}^{\lambda}_{\hat{\gamma}}\widetilde{e}^{\hat{\beta}\mu}\right)d^4 x$$
$$+ (a+b)\int \sqrt{-g}(\Lambda^{\hat{\alpha}}_{\hat{\beta},\rho}\overline{\Lambda}^{\hat{\gamma}}_{\hat{\alpha},\sigma}\widetilde{e}^{\rho}_{\hat{\gamma}}\widetilde{e}^{\hat{\beta}\sigma} - 2\Lambda^{\hat{\alpha}}_{\hat{\beta}}\overline{\Lambda}^{\hat{\gamma}}_{\hat{\alpha},\rho}\widetilde{e}^{\rho}_{\hat{\gamma},\sigma}\widetilde{e}^{\hat{\beta}\sigma})d^4 x + \varepsilon \int \Delta_\varepsilon d^4 x.$$

We see that if $a+b=0, \varepsilon=0$, then the change of the action $L$ given by (1-18) under local Lorentz transformation is just only an integral of a total derivative. However, this is just the case of (1-17) with the cosmological constant $\kappa$. In this paper we don't take account of the cosmological constant, if we must take account of it, then what we have to make is just only adding $\kappa$ into (1-17). We therefore have proved that the change of $L_G$ given by (1-17) under local Lorentz transformation is only to raise an integral of a total derivative in the Einstein-Hilbert action, which does not impact on the derivation of the equations of motion. On the other hand, the above result also shows that why the term $F^{\mu\nu\lambda}F_{\mu\nu\lambda}$ cannot appears in $L_G$.

Under local Lorentz transformation, assuming that the manner of transformation of the wave function $\Psi(x)$ in (1-11) reads:

$$\Psi(x) \to S(x)\Psi(x),$$

where $S(x)$ satisfies

$$S(x)\Lambda^{\hat{\alpha}}_{\hat{\beta}}(x)\gamma^{\hat{\beta}}S^{-1}(x) = \gamma^{\hat{\alpha}},$$

we can prove that such $S(x)$ satisfies

$$iS^{-1}(x)\frac{\partial S(x)}{\partial x^\mu} + \frac{1}{4}\Lambda^{\hat{\alpha}}_{\hat{\beta}}(x)\frac{\partial \overline{\Lambda}^{\hat{\gamma}}_{\hat{\alpha}}(x)}{\partial x^\mu}\eta_{\hat{\gamma}\hat{\chi}}\sigma^{\hat{\beta}\hat{\chi}} = 0.$$

Combining above formula and (1-20) we can prove that both (1-10) and (1-11) are invariant under local Lorentz transformation.

### 1.3 The Einstein equations

For the sake of brevity, we define

$$S^{\mu\nu\lambda} = F^{\mu\nu\lambda} + g^{\lambda\mu} F^{\nu\sigma}{}_\sigma - g^{\lambda\nu} F^{\mu\sigma}{}_\sigma ,$$

from the above formula we have

$$S^{\mu\lambda}{}_\lambda = -2F^{\mu\lambda}{}_\lambda ;$$

And, thus, we have

$$F^{\mu\nu\lambda} = S^{\mu\nu\lambda} + \frac{1}{2} g^{\lambda\mu} S^{\nu\sigma}{}_\sigma - \frac{1}{2} g^{\lambda\nu} S^{\mu\sigma}{}_\sigma .$$

According to (1-14) and (1-15) we can prove the Einstein tensor

$$\begin{aligned} R^{\mu\nu} - \frac{1}{2} g^{\mu\nu} R &= S^{\mu\lambda\nu}{}_{;\lambda} + F^{\rho\mu\nu} F_{\rho\sigma}{}^\sigma - F^{\rho\mu\sigma} F_{\rho\sigma}{}^\nu - \frac{1}{2} g^{\mu\nu} (F^{\rho\sigma}{}_\sigma F_{\rho\lambda}{}^\lambda - F^{\rho\sigma\lambda} F_{\rho\lambda\sigma}) \\ &= \frac{1}{\sqrt{-g}} \frac{\partial}{\partial x^\lambda} \left( \sqrt{-g} S^{\mu\lambda\nu} \right) + W^{\mu\nu} , \end{aligned} \quad (1\text{-}21)$$

where

$$W^{\mu\nu} = \Gamma^\nu_{\rho\sigma} S^{\mu\rho\sigma} + S^{\mu\rho\sigma} F_{\rho\sigma}{}^\nu - \frac{1}{2} g^{\mu\nu} L_G , \quad (1\text{-}22)$$

$L_G$ is given by (1-17), for the Christoffel symbol $\Gamma^\nu_{\rho\sigma} = \frac{1}{2} g^{\nu\mu} (g_{\mu\rho,\sigma} + g_{\mu\sigma,\rho} - g_{\rho\sigma,\mu})$ in (1-22), according to (1-5) we can prove

$$\Gamma^\nu_{\rho\sigma} = e^\nu_{\hat{\alpha}} e^{\hat{\alpha}}_{\rho,\sigma} - F^\nu{}_{\rho\sigma} .$$

Based on (1-21), the Einstein equations $R^{\mu\nu} - \frac{1}{2} g^{\mu\nu} R = \frac{8\pi G}{c^4} T^{\mu\nu}$ can be written to the following two equivalent forms:

$$S^{\mu\lambda\nu}{}_{;\lambda} + S^{\mu\rho\sigma} F_{\rho\sigma}{}^\nu - \frac{1}{2} g^{\mu\nu} L_G = \frac{8\pi G}{c^4} T^{\mu\nu} ,$$

$$\frac{\partial}{\partial x^\lambda} \left( \sqrt{-g} S^{\mu\lambda\nu} \right) + \sqrt{-g} W^{\mu\nu} = \frac{8\pi G}{c^4} \sqrt{-g} T^{\mu\nu} .$$

According to the above second equivalent form of the Einstein equations and notice $S^{\mu\nu\lambda} = -S^{\nu\mu\lambda}$ we have

$$\frac{\partial}{\partial x^\mu} \left[ \sqrt{-g} \left( T^{\mu\nu} - \frac{c^4}{8\pi G} W^{\mu\nu} \right) \right] = \frac{c^4}{8\pi G} \frac{\partial}{\partial x^\mu} \frac{\partial}{\partial x^\lambda} \left( \sqrt{-g} S^{\mu\lambda\nu} \right) = 0 .$$

Defining

$$\begin{aligned} S^{\mu\nu\hat{\alpha}} &= S^{\mu\nu\lambda} e^{\hat{\alpha}}_\lambda \\ &= \frac{1}{2} \left( e^\mu_{\hat{\beta}} g^{\nu\rho} - e^\nu_{\hat{\beta}} g^{\mu\rho} \right) e^{\hat{\alpha}\sigma} \left( e^{\hat{\beta}}_{\rho,\sigma} - e^{\hat{\beta}}_{\sigma,\rho} \right) \\ &\quad - \left( e^{\hat{\alpha}\mu} g^{\nu\rho} - e^{\hat{\alpha}\nu} g^{\mu\rho} \right) e^\sigma_{\hat{\beta}} \left( e^{\hat{\beta}}_{\rho,\sigma} - e^{\hat{\beta}}_{\sigma,\rho} \right) - \frac{1}{2} g^{\mu\rho} g^{\nu\sigma} \left( e^{\hat{\alpha}}_{\rho,\sigma} - e^{\hat{\alpha}}_{\sigma,\rho} \right) , \end{aligned} \quad (1\text{-}23)$$

$$\Theta^{\mu\hat{\alpha}} = \frac{1}{|^4 e|} \frac{\partial}{\partial x^\nu} \left( |^4 e| S^{\mu\nu\hat{\alpha}} \right) + e^{\hat{\alpha}\rho} S^{\mu\sigma\hat{\beta}} \left( e_{\hat{\beta}\rho,\sigma} - e_{\hat{\beta}\sigma,\rho} \right) - \frac{1}{2} e^{\hat{\alpha}\mu} L_G - \frac{8\pi G}{c^4} T^{\mu\hat{\alpha}} , \quad (1\text{-}24)$$

where $T^{\mu\hat{\alpha}} = T^{\mu\nu} e^{\hat{\alpha}}_\nu$, the Einstein equations can be written to the following form

$$\Theta^{\mu\hat{\alpha}} = 0 . \quad (1\text{-}25)$$

According to the above form of the Einstein equations and notice $S^{\mu\nu\hat{\alpha}} = -S^{\nu\mu\hat{\alpha}}$ we have

$$\frac{\partial}{\partial x^{\mu}}\left(|^{4}e|\left\{T^{\mu\hat{\alpha}} + \frac{c^{4}}{8\pi G}\left[\frac{1}{2}e^{\hat{\alpha}\mu}L_{G} - e^{\hat{\alpha}\rho}S^{\mu\sigma\hat{\beta}}\left(e_{\hat{\beta}\rho,\sigma} - e_{\hat{\beta}\sigma,\rho}\right)\right]\right\}\right) = \frac{c^{4}}{8\pi G}\frac{\partial}{\partial x^{\mu}}\frac{\partial}{\partial x^{\nu}}\left(\sqrt{-g}S^{\mu\nu\hat{\alpha}}\right) = 0.$$

In fact, (1-25) is just the Euler-Lagrange equations

$$\frac{\partial\left[|^{4}e|\left(\frac{c^{3}}{16\pi G}L_{G} + L_{M}\right)\right]}{\partial e_{\hat{\alpha}\mu}} - \partial_{\nu}\frac{\partial\left[|^{4}e|\left(\frac{c^{3}}{16\pi G}L_{G} + L_{M}\right)\right]}{\partial e_{\hat{\alpha}\mu,\nu}} = -\frac{c^{3}|^{4}e|}{8\pi G}\Theta^{\mu\hat{\alpha}} = 0,$$

where $L_{M}$ is the Lagrangian of matter, e.g., for the Dirac field, $L_{M} = L_{D}$, $L_{D}$ is given by (1-11);

$$T^{\mu\hat{\alpha}} = \frac{c}{|^{4}e|}\left[\frac{\partial\left(|^{4}e|L_{M}\right)}{\partial e_{\hat{\alpha}\mu}} - \partial_{\nu}\frac{\partial\left(|^{4}e|L_{M}\right)}{\partial e_{\hat{\alpha}\mu,\nu}}\right].$$

Although there are the sixteen Euler-Lagrange equations in (1-25), defining

$$S^{\mu\lambda\nu}_{(\pm)} = \frac{1}{2}(S^{\mu\lambda\nu} \pm S^{\nu\lambda\mu}), \quad W^{\mu\nu}_{(\pm)} = \frac{1}{2}(W^{\mu\nu} \pm W^{\nu\mu}),$$

and using (1-21) we can obtain the six identities

$$\frac{\partial}{\partial x^{\lambda}}\left(\sqrt{-g}S^{\mu\lambda\nu}_{(-)}\right) + \frac{1}{2}\sqrt{-g}W^{\mu\nu}_{(-)} = 0 \qquad (1\text{-}26)$$

we therefore have only the ten independent equations

$$\frac{\partial}{\partial x^{\lambda}}\left(\sqrt{-g}S^{\mu\lambda\nu}_{(+)}\right) + \frac{1}{2}\sqrt{-g}W^{\mu\nu}_{(+)} = \frac{8\pi G}{c^{4}}\sqrt{-g}T^{\mu\nu}.$$

Using (1-26) we can prove

$$\Theta^{\mu\nu} = \Theta^{\nu\mu} = \frac{1}{\sqrt{-g}}\frac{\partial}{\partial x^{\lambda}}\left(\sqrt{-g}S^{\mu\lambda\nu}_{(+)}\right) + W^{\mu\nu}_{(+)} - \frac{8\pi G}{c^{4}}T^{\mu\nu}. \qquad (1\text{-}27)$$

Notice that in $L_{G}$ given by (1-17) and all the forms of the Einstein equations, the basic variables what we use are tetrad $\{e^{\hat{\alpha}}_{\mu}\}$, $g_{\mu\nu}$ and $g^{\mu\nu}$ are as the abbreviation for $e^{\hat{\alpha}}_{\mu}e_{\hat{\alpha}\nu}$ and $e^{\hat{\alpha}\mu}e^{\nu}_{\hat{\alpha}}$, respectively. Thus, both the Einstein-Hilbert action and the Einstein equations are expressed absolutely by tetrad $\{e^{\hat{\alpha}}_{\mu}\}$.

## 2  The Schwinger time gauge condition

The invariance of the form of the theory given by §1 under coordinate transformation $x^{\mu} = x^{\mu}(\tilde{x}^{\nu})$ and local Lorentz transformation indicated by (1-19) implies the existence of ten arbitrary gauge functions, which must be eliminated by adding ten gauge conditions to 16 variables $e^{\hat{\alpha}}_{\mu}$ in project of canonical quantization. Among ten gauge conditions, the four are for coordinate transformation and the six are for local Lorentz transformation. For avoiding the influence between the two kinds of transformation as far as possible, we restrict that coordinate transformation $x^{\mu} = x^{\mu}(\tilde{x}^{\nu})$ is only used to determine a special group of $g_{\mu\nu}$ but not to determine $e^{\hat{\alpha}}_{\mu}$ directly, it acts on $e^{\hat{\alpha}}_{\mu}$ through (1-5). Of course, this rule cannot avoid the influence absolutely.

Although that gauge-fixing term is not destined at beginning may be a better method for the discussion, for the sake of simpleness, we first choose three gauge-fixing terms:

$$e_{\hat{a}}^0 = 0, \quad a = 1, 2, 3, \qquad (2\text{-}1)$$

which can be implemented by choosing appropriate $\Lambda_{\hat{0}}^{\hat{a}}(x)$ of $\Lambda_{\hat{\beta}}^{\hat{\alpha}}(x)$ indicated by (1-19). It then leads to the following results:

$$e_i^{\hat{0}} = 0, \ i = 1,2,3; \ e_{\hat{0}}^0 = \left(e_0^{\hat{0}}\right)^{-1}; \ e_{\hat{0}}^i = -e_{\hat{0}}^0 e_0^{\hat{a}} e_{\hat{a}}^i; \ \sqrt{-g} = \left|{}^4e\right| = e_0^{\hat{0}}\left|{}^3e\right|; \ e_{\hat{a}}^i e_i^{\hat{b}} = \delta_{\hat{a}}^{\hat{b}}; \ e_{\hat{a}}^j e_i = \delta_i^j, \ (2\text{-}2)$$

where $\left|{}^3e\right| = \det\left[e_i^{\hat{a}}\right]$ is the determinant of the $3 \times 3$ matrix $\left[e_i^{\hat{a}}\right]$.

The condition (2-1), so called "time gauge", was first given by J. Schwinger[1]; its equivalent form $e_i^{\hat{0}} = 0$ is introduced by other consideration recently[6].

Under the condition (2-1), (1-17) is simplified to the following form

$$L_G = L_{G0} + 2e_0^0 e_{0,k}^{\hat{0}} U^k + L_{GV},$$

$$L_{G0} = \frac{1}{2} M_{\hat{a}\hat{b}}^{ij} e_{\hat{0}}^{\mu}\left(e_{i,\mu}^{\hat{a}} - e_{\mu,i}^{\hat{a}}\right) e_{\hat{0}}^{\nu}\left(e_{j,\nu}^{\hat{b}} - e_{\nu,j}^{\hat{b}}\right) = \left(e_0^0\right)^2 \overline{L}_{G0},$$

$$U^k = e_{\hat{a}}^i e_{\hat{b}}^j e^{\hat{b}k}\left(e_{i,j}^{\hat{a}} - e_{j,i}^{\hat{a}}\right),$$

$$L_{GV} = e^{\hat{c}l} e_{\hat{c}}^m \left(-\frac{1}{4}\eta_{\hat{a}\hat{b}} e^{\hat{d}i} e_{\hat{d}}^j - \frac{1}{2} e_{\hat{a}}^j e_{\hat{b}}^i + e_{\hat{a}}^i e_{\hat{b}}^j\right)\left(e_{i,l}^{\hat{a}} - e_{l,i}^{\hat{a}}\right)\left(e_{j,m}^{\hat{b}} - e_{m,j}^{\hat{b}}\right), \qquad (2\text{-}3)$$

$$M_{\hat{a}\hat{b}}^{ij} = \eta_{\hat{a}\hat{b}} e^{\hat{c}i} e_{\hat{c}}^j + e_{\hat{a}}^j e_{\hat{b}}^i - 2 e_{\hat{a}}^i e_{\hat{b}}^j,$$

$$\overline{L}_{G0} = \frac{1}{2} M_{\hat{a}\hat{b}}^{ij}\left(e_{i,0}^{\hat{a}} - e_{0,i}^{\hat{a}}\right)\left(e_{j,0}^{\hat{b}} - e_{0,j}^{\hat{b}}\right) - M_{\hat{a}\hat{b}}^{ij}\left(e_{i,0}^{\hat{a}} - e_{0,i}^{\hat{a}}\right) e_0^{\hat{d}} e_{\hat{d}}^m\left(e_{j,m}^{\hat{b}} - e_{m,j}^{\hat{b}}\right)$$

$$+ \frac{1}{2} M_{\hat{a}\hat{b}}^{ij} e_0^{\hat{c}} e_{\hat{c}}^l\left(e_{i,l}^{\hat{a}} - e_{l,i}^{\hat{a}}\right) e_0^{\hat{d}} e_{\hat{d}}^m\left(e_{j,m}^{\hat{b}} - e_{m,j}^{\hat{b}}\right).$$

Notice that there is not $e_{\hat{0}}^0$ or $e_0^{\hat{0}}$ in $\overline{L}_{G0}$, time derivative term only appears in $L_{G0}$ and there is not the term $e_{0,0}^{\hat{a}}$ in $L_{G0}$. In fact, there is not the term $e_{\lambda,\lambda}^{\hat{\alpha}}$ ($\alpha, \lambda = 0, 1, 2, 3$) in $L_G$ given by (1-17).

We can prove that the determinant of the $9 \times 9$ symmetric matrix $M_{\hat{a}\hat{b}}^{ij}$

$$\left|M_{\hat{a}\hat{b}}^{ij}\right| = 0;$$

And, further, the equation of eigenvalues of $M_{\hat{a}\hat{b}}^{ij}$ is

$$\left|M_{\hat{a}\hat{b}}^{ij} - \lambda I\right| = -\lambda^3 f_1(\lambda) f_2(\lambda) = 0,$$

$$f_1(\lambda) = \lambda^3 - \frac{4}{\left|{}^3e\right|^2}\left[\sum_{a,i=1}^{3}\left(e_i^{\hat{a}}\right)^2\right]\lambda + \frac{16}{\left|{}^3e\right|^2},$$

$$f_2(\lambda) = \lambda^3 - 2\left[\sum_{a,i=1}^{3}\left(e_{\hat{a}}^i\right)^2\right]\lambda^2 + \left[\sum_{a,i=1}^{3}\left(e_{\hat{a}}^i\right)^2 + \frac{1}{\left|{}^3e\right|^2}\sum_{a,i=1}^{3}\left(e_i^{\hat{a}}\right)^2\right]\lambda - \frac{1}{\left|{}^3e\right|^2}\left[\sum_{a,i=1}^{3}\left(e_i^{\hat{a}}\right)^2\sum_{a,i=1}^{3}\left(e_{\hat{a}}^i\right)^2 - 1\right].$$

We see that the rank of the 9×9 symmetric matrix $M_{\hat{a}\hat{b}}^{ij}$ is 6. This result is foreseeable because there are still three undecided spatial elements of rotation in local Lorentz transformation $\Lambda_{\hat{\beta}}^{\hat{\alpha}}(x)$, hence, there are three arbitrary gauge functions for $\Lambda_{\hat{b}}^{\hat{a}}(x)$.

Although we have only the ten independent Einstein equations, we analyze generally the sixteen Euler-Lagrange equations $\Theta^{\mu\hat{\alpha}} = 0$ under the condition (2-1). At first, the four equations of constraint in which there is not any second time derivative term read

$$-2e_0^{\hat{0}}\Theta^{00} = -2\Theta^{\hat{0}\hat{0}} = L_{G0} + \frac{2}{\left|{}^3e\right|}\frac{\partial}{\partial x^k}\left(\left|{}^3e\right|U^k\right) - L_{GV} + \frac{16\pi G}{c^4}T^{\hat{0}\hat{0}} = 0, \qquad (2\text{-}4)$$

$$e_0^{\hat{0}}\Theta^{0\hat{a}} = \Theta^{\hat{0}\hat{a}} = \frac{1}{\left|{}^3e\right|}\frac{\partial}{\partial x^k}\left(\left|{}^3e\right|S^{\hat{0}k\hat{a}}\right) + e^{\hat{a}i}S^{\hat{0}j\hat{b}}\left(e_{\hat{b}i,j} - e_{\hat{b}j,i}\right) - \frac{8\pi G}{c^4}T^{\hat{0}\hat{a}} = 0, \qquad (2\text{-}5)$$

where

$$S^{\hat{0}i\hat{a}} = e_0^{\hat{0}}S^{0i\hat{a}} = e^{\hat{a}i}e_{\hat{0}}^{\lambda}e_{\hat{b}}^{j}\left(e_{j,\lambda}^{\hat{b}} - e_{\lambda,j}^{\hat{b}}\right) - \frac{1}{2}e_{\hat{b}}^{i}e_{\hat{0}}^{\lambda}\left[e^{\hat{b}j}\left(e_{j,\lambda}^{\hat{a}} - e_{\lambda,j}^{\hat{a}}\right) + e^{\hat{a}j}\left(e_{j,\lambda}^{\hat{b}} - e_{\lambda,j}^{\hat{b}}\right)\right]. \qquad (2\text{-}6)$$

Notice that (2-4) leads to

$$\left(e_{\hat{0}}^0\right)^2 = \left(e_0^{\hat{0}}\right)^{-2} = \frac{-\frac{2}{\left|{}^3e\right|}\frac{\partial}{\partial x^k}\left(\left|{}^3e\right|U^k\right) + L_{GV} - \frac{16\pi G}{c^4}T^{\hat{0}\hat{0}}}{\overline{L}_{G0}} \geq 0. \qquad (2\text{-}7)$$

Because there is not $e_{\hat{0}i}$ in $e_0^{\hat{0}}\left|{}^3e\right|L_G$, there is not the corresponding equation $\Theta^{i\hat{0}} = 0$ in the Euler-Lagrange equations. On the other hand, because $\Theta^{0\hat{a}} = \Theta^{0\mu}e_{\mu}^{\hat{a}} = \Theta^{00}e_0^{\hat{a}} + \Theta^{0i}e_i^{\hat{a}}$, according to (1-27) and (2-2) we have $\Theta^{i\hat{0}} = \Theta^{i\mu}e_{\mu}^{\hat{0}} = \Theta^{i0}e_0^{\hat{0}} = \Theta^{0i}e_0^{\hat{0}} = e_{\hat{a}}^i\left(\Theta^{0\hat{a}}e_0^{\hat{0}} - \Theta^{00}e_0^{\hat{a}}\right)$, we see that $\Theta^{i\hat{0}} = 0$ does not provide new independent equation.

The rest nine equations are

$$e_{\hat{0}}^{\hat{0}} \Theta^{i\hat{a}} = -\frac{1}{|{}^3 e|} \frac{\partial}{\partial x^0} \left( |{}^3 e| S^{\hat{0}i\hat{a}} \right) + \frac{1}{|{}^3 e|} \frac{\partial}{\partial x^j} \left[ |{}^3 e| \left( e_0^{\hat{b}} e_{\hat{b}}^j S^{\hat{0}i\hat{a}} - e_0^{\hat{b}} e_{\hat{b}}^i S^{\hat{0}j\hat{a}} + e_0^{\hat{0}} \tilde{s}^{ij\hat{a}} \right) \right]$$
$$+ e_{0,j}^{\hat{0}} e^{\hat{a}j} U^i - e^{\hat{a}j} S^{\hat{0}i\hat{b}} \left( e_{\hat{b}j,0} - e_{\hat{b}0,j} \right) \qquad (2\text{-}8)$$
$$+ e^{\hat{a}j} \left( e_0^{\hat{c}} e_{\hat{c}}^k S^{\hat{0}i\hat{b}} - e_0^{\hat{0}} e_{\hat{c}}^i S^{\hat{0}k\hat{b}} + e_0^{\hat{0}} \tilde{s}^{ik\hat{b}} \right) \left( e_{\hat{b}j,k} - e_{\hat{b}k,j} \right) - \frac{8\pi G}{c^4} e_0^{\hat{0}} T^{i\hat{a}} = 0,$$

where $U^i$ and $S^{\hat{0}i\hat{a}}$ are given by (2-3) and (2-6), respectively;

$$\tilde{s}^{ij\hat{a}} = e_{\hat{0}}^0 e_{0,k}^{\hat{0}} e_{\hat{b}}^k \left( e^{\hat{a}i} e^{\hat{b}j} - e^{\hat{b}i} e^{\hat{a}j} \right) - \frac{1}{2} e_{\hat{b}}^i e^{\hat{b}l} e_{\hat{c}}^j e^{\hat{c}m} \left( e_{l,m}^{\hat{a}} - e_{m,l}^{\hat{a}} \right)$$
$$+ \frac{1}{2} e^{\hat{a}l} e^{\hat{b}m} \left( e_{\hat{b}}^i e_{\hat{c}}^j - e_{\hat{b}}^j e_{\hat{c}}^i \right) \left( e_{l,m}^{\hat{c}} - e_{m,l}^{\hat{c}} \right) - \left( e^{\hat{a}i} e^{\hat{b}j} - e^{\hat{b}i} e^{\hat{a}j} \right) e_{\hat{b}}^l e_{\hat{c}}^m \left( e_{l,m}^{\hat{c}} - e_{m,l}^{\hat{c}} \right). \qquad (2\text{-}9)$$

## 3  Non-positive definiteness of the quadratic term of time derivative in $L_G$

### 3.1  Non-positive definiteness of the quadratic term of time derivative in $L_G$

There exists a basic problem in $L_G$ given by (2-3): the quadratic term of time derivative in $L_{G0}$ is non-positive definitive. This conclusion is obvious from the following expression:

$$L_{G0} = -\frac{2}{3} \left[ e_{\hat{0}}^\lambda e_{\hat{a}}^i \left( e_{i,\lambda}^{\hat{a}} - e_{\lambda,i}^{\hat{a}} \right) \right]^2 + \frac{1}{6} \left\{ e_{\hat{0}}^\lambda \left[ 2e^{\hat{1}i} \left( e_{i,\lambda}^{\hat{1}} - e_{\lambda,i}^{\hat{1}} \right) - e^{\hat{2}i} \left( e_{i,\lambda}^{\hat{2}} - e_{\lambda,i}^{\hat{2}} \right) - e^{\hat{3}i} \left( e_{i,\lambda}^{\hat{3}} - e_{\lambda,i}^{\hat{3}} \right) \right] \right\}^2$$
$$+ \frac{1}{2} \left\{ e_{\hat{0}}^\lambda \left[ e^{\hat{2}i} \left( e_{i,\lambda}^{\hat{2}} - e_{\lambda,i}^{\hat{2}} \right) - e^{\hat{3}i} \left( e_{i,\lambda}^{\hat{3}} - e_{\lambda,i}^{\hat{3}} \right) \right] \right\}^2 + \frac{1}{2} \left\{ e_{\hat{0}}^\lambda \left[ e^{\hat{1}i} \left( e_{i,\lambda}^{\hat{2}} - e_{\lambda,i}^{\hat{2}} \right) + e^{\hat{2}i} \left( e_{i,\lambda}^{\hat{1}} - e_{\lambda,i}^{\hat{1}} \right) \right] \right\}^2 \qquad (3\text{-}1)$$
$$+ \frac{1}{2} \left\{ e_{\hat{0}}^\lambda \left[ e^{\hat{1}i} \left( e_{i,\lambda}^{\hat{3}} - e_{\lambda,i}^{\hat{3}} \right) + e^{\hat{3}i} \left( e_{i,\lambda}^{\hat{1}} - e_{\lambda,i}^{\hat{1}} \right) \right] \right\}^2 + \frac{1}{2} \left\{ e_{\hat{0}}^\lambda \left[ e^{\hat{2}i} \left( e_{i,\lambda}^{\hat{3}} - e_{\lambda,i}^{\hat{3}} \right) + e^{\hat{3}i} \left( e_{i,\lambda}^{\hat{2}} - e_{\lambda,i}^{\hat{2}} \right) \right] \right\}^2 .$$

The six terms in (3-1) are independent each other, because we have proved that the rank of the $9 \times 9$ symmetric matrix $M_{\hat{a}\hat{b}}^{ij}$ in $L_{G0}$ is 6. However, the reason that we obtain the conclusion "the quadratic term of time derivative in $L_{G0}$ is non-positive definitive" is that we have taken advantage of the condition (2-1), hence, a question is whether this conclusion holds in general case, namely, does it hold for the action (1-1)? We discuss this question as follows.

Using (2-1) and (2-2), from (1-5) we have

$$g^{00} = -\left( e_{\hat{0}}^0 \right)^2 , \quad g^{0i} = -e_{\hat{0}}^0 e_{\hat{0}}^i , \quad g^{ij} = -e_{\hat{0}}^i e_{\hat{0}}^j + e_{\hat{a}}^i e^{\hat{a}j} ;$$
$$g_{00} = -\left( e_0^{\hat{0}} \right)^2 + e_0^{\hat{a}} e_{\hat{a}0} , \quad g_{0i} = e_0^{\hat{a}} e_{\hat{a}i} , \quad g_{ij} = e_i^{\hat{a}} e_{\hat{a}j} . \qquad (3\text{-}2)$$

From (1-5) we have $e_{\hat{\alpha}}^\mu e_{\hat{\beta}}^\nu g_{\mu\nu,\lambda} = e_{\hat{\alpha}}^\mu e_{\hat{\beta}\mu,\lambda} + e_{\hat{\beta}}^\mu e_{\hat{\alpha}\mu,\lambda}$, using this formula and considering (2-1), (2-2) and (3-2), we can prove

$$e_{\hat{0}}^\lambda e_{\hat{a}}^i \left( e_{\hat{b}i,\lambda} - e_{\hat{b}\lambda,i} \right) + e_{\hat{0}}^\lambda e_{\hat{b}}^i \left( e_{\hat{a}i,\lambda} - e_{\hat{a}\lambda,i} \right) = 2 e_{\hat{0}}^\lambda e_{\hat{a}}^i e_{\hat{b}}^j \Gamma_{\lambda ij} ,$$

where $\Gamma_{\lambda ij} = \frac{1}{2} \left( g_{\lambda i,j} + g_{\lambda j,i} - g_{ij,\lambda} \right)$. We therefore have

$$L_{G0} = -\frac{2}{3}\left(e_{\hat{0}}^{\lambda}e^{\hat{a}i}e_{\hat{a}}^{j}\Gamma_{\lambda ij}\right)^2 + \frac{1}{6}\left[e_{\hat{0}}^{\lambda}\left(2e_{\hat{1}}^{i}e_{\hat{1}}^{j} - e_{\hat{2}}^{i}e_{\hat{2}}^{j} - e_{\hat{3}}^{i}e_{\hat{3}}^{j}\right)\Gamma_{\lambda ij}\right]^2 + \frac{1}{2}\left[e_{\hat{0}}^{\lambda}\left(e_{\hat{2}}^{i}e_{\hat{2}}^{j} - e_{\hat{3}}^{i}e_{\hat{3}}^{j}\right)\Gamma_{\lambda ij}\right]^2$$

$$+ 2\left(e_{\hat{0}}^{\lambda}e_{\hat{1}}^{i}e_{\hat{2}}^{j}\Gamma_{\lambda ij}\right)^2 + 2\left(e_{\hat{0}}^{\lambda}e_{\hat{1}}^{i}e_{\hat{3}}^{j}\Gamma_{\lambda ij}\right)^2 + 2\left(e_{\hat{0}}^{\lambda}e_{\hat{2}}^{i}e_{\hat{3}}^{j}\Gamma_{\lambda ij}\right)^2 \quad (3\text{-}3)$$

$$= e_{\hat{0}}^{\mu}e_{\hat{0}}^{\nu}\left(\overline{g}^{il}\overline{g}^{jm} - \overline{g}^{ij}\overline{g}^{lm}\right)\Gamma_{\mu ij}\Gamma_{\nu lm},$$

where $\overline{g}^{ij} = e_{\hat{a}}^{i}e^{\hat{a}j} = g^{ij} + e_{\hat{0}}^{i}e_{\hat{0}}^{j}$, and $\overline{g}^{ik}g_{kj} = \delta_{j}^{i}$.

In (3-3), the quadratic term of time derivative $L_{G00}$ is

$$L_{G00} = \frac{1}{2}M_{\hat{a}\hat{b}}^{ij}e_{i,0}^{\hat{a}}e_{j,0}^{\hat{b}}$$

$$= \frac{1}{4}\left(e_{\hat{0}}^{0}\right)^2\left(\overline{g}^{il}\overline{g}^{jm} - \overline{g}^{ij}\overline{g}^{lm}\right)g_{ij,0}g_{lm,0}$$

$$= \frac{1}{2}\frac{1}{-g}\left[g_{11}(g_{23,0})^2 + g_{22}(g_{31,0})^2 + g_{33}(g_{12,0})^2\right. \quad (3\text{-}4)$$

$$- g_{11}g_{22,0}g_{33,0} - g_{22}g_{33,0}g_{11,0} - g_{33}g_{11,0}g_{22,0}$$

$$+ 2g_{12}g_{12,0}g_{33,0} + 2g_{23}g_{23,0}g_{11,0} + 2g_{31}g_{31,0}g_{22,0}$$

$$\left. - g_{12}g_{23,0}g_{31,0} - g_{23}g_{31,0}g_{12,0} - g_{31}g_{12,0}g_{23,0}\right].$$

Using (3-2), we can prove that all the three principal minors of the metric $g_{ij}$ are positive, e.g.,

$$g_{33} = \sum_{a=1}^{3}\left(e_{3}^{\hat{a}}\right)^2 > 0, \quad \begin{vmatrix} g_{22} & g_{23} \\ g_{32} & g_{33} \end{vmatrix} = \left(e_{2}^{\hat{1}}e_{3}^{\hat{2}} - e_{3}^{\hat{1}}e_{2}^{\hat{2}}\right)^2 + \left(e_{2}^{\hat{1}}e_{3}^{\hat{3}} - e_{3}^{\hat{1}}e_{2}^{\hat{3}}\right)^2 + \left(e_{2}^{\hat{2}}e_{3}^{\hat{3}} - e_{3}^{\hat{2}}e_{2}^{\hat{3}}\right)^2 > 0, \quad (3\text{-}5)$$

$$|g_{ij}| = |^{3}e|^2 > 0.$$

Considering (3-5), we can introduce a group of new variables $h_u, u = 0, 1, 2, 3, 4, 5$:

$$\sqrt{|g_{ij}|} = h_0, \quad \frac{\sqrt[4]{\left[g_{22}g_{33} - (g_{23})^2\right]^3}}{\sqrt{|g_{ij}|}} = h_1, \quad \frac{\sqrt{g_{33}}}{\sqrt[4]{g_{22}g_{33} - (g_{23})^2}} = h_2,$$

$$h_3 = \frac{g_{23}}{g_{33}}, \quad h_4 = \frac{g_{23}g_{31} - g_{12}g_{33}}{g_{22}g_{33} - (g_{23})^2}, \quad h_5 = \frac{g_{12}g_{23} - g_{22}g_{31}}{g_{22}g_{33} - (g_{23})^2}; \quad (3\text{-}6)$$

Contrarily, according to (3-6) we can obtain easily the expression

$$g_{ij} = g_{ij}(h_u), u = 0, 1, 2, 3, 4, 5 \quad (3\text{-}7)$$

And, from (3-6) we see that if there is not gravitation field, and $g_{11} = g_{22} = g_{33} = 1$, $g_{12} = g_{23} = g_{31} = 0$ (i.e., Minkowski metric), then $h_0 = h_1 = h_2 = 1, h_3 = h_4 = h_5 = 0$.

Using (3-6) we can prove

$$L_{G00} = \left(e_{\hat{0}}^{0}\right)^2\left[-\frac{2}{3}\frac{1}{h_0^2}(h_{0,0})^2 + \frac{2}{3}\frac{1}{h_1^2}(h_{1,0})^2 + \frac{2}{h_2^2}(h_{2,0})^2\right.$$

$$\left. + \frac{1}{2}h_2^4(h_{3,0})^2 + \frac{1}{2}\frac{h_1^2}{h_2^2}(h_{4,0})^2 + \frac{1}{2}h_1^2h_2^2(h_3h_{4,0} + h_{5,0})^2\right]. \quad (3\text{-}8)$$

We see that, taking advantage of the group of variables substitution (3-6), $\sqrt{|g_{ij}|}$ as an independent variable is separated from the six dynamical variables $g_{ij}$.

The result (3-8) shows again that the six terms in (3-1) are independent each other and shows clearly that the quadratic term of time derivative in $L_{G0}$ is non-positive definitive.

In general case, the Lagrangian $L_g$ is given by (1-2) and we can prove that the quadratic term of time derivative in $L_g$ is

$$L_{g0} = \frac{1}{4}\left[-2g^{0i}\left(g^{0l}g^{jm} - g^{0j}g^{lm}\right) + g^{00}\left(g^{ij}g^{lm} - g^{il}g^{jm}\right)\right]g_{ij,0}g_{lm,0}$$

$$= \frac{1}{4}\left(-g^{00}\right)\left(\tilde{g}^{il}\tilde{g}^{jm} - \tilde{g}^{ij}\tilde{g}^{lm}\right)g_{ij,0}g_{lm,0} .$$

where $\tilde{g}^{ij} = g^{ij} - \dfrac{g^{0i}g^{0j}}{g^{00}}$, and $\tilde{g}^{ik}g_{kj} = \delta^i_j$.

However, in this case, the characteristics indicated by (3-5) do not hold for metric $g_{ij}$ and thus, generally speaking, we cannot judge whether $L_{g0}$ is positive definitive.

On the other hand, for the general case, if we *assume* that the characteristics indicated by (3-5) hold (This is just physically significant case), then we can still obtain (3-8) by (3-6) for $L_{g0}$. This discussion shows that the characteristic that the quadratic term of time derivative in the Einstein-Hilbert action is non-positive definitive is ineluctable.

In my opinion, the quadratic term of time derivative in an action should be positive definitive. Because, the quadratic term of time derivative in an action corresponds to the kinetic energy of the system, if this term was non-positive, then it was weird. On the other hand, the non-positive definiteness of the quadratic term of time derivative in an action leads to the principle of variation failure[10].

However, we emphasize that the characteristic "the quadratic term of time derivative in the Einstein-Hilbert action is non-positive definitive" does not denote that there is inconsistency in the structure of the theory of general relativity, it only shows that if we regard general relativity as a theory of field (e.g. tetrad field), then this characteristic is incongruous with theory of field.

On the other hand, if we want to obtain a positive definitive kinetic energy in the Lagrangian of the Einstein-Hilbert action, it is obvious that unique method is that we choose some "gauge-fixing term" such that there is not any time derivative term in the first term of (3-1), i.e., the term $-\dfrac{2}{3}\left[e^{\lambda}_0 e^i_{\hat{a}}\left(e^{\hat{a}}_{i,\lambda} - e^{\hat{a}}_{\lambda,i}\right)\right]^2$; And no matter how we choose "gauge-fixing term", the necessary condition is that the term $e^i_{\hat{a}} e^{\hat{a}}_{i,0} = \dfrac{1}{\left|{}^3e\right|}\dfrac{\partial\left|{}^3e\right|}{\partial x^0} = \dfrac{1}{2}\dfrac{1}{\left|g_{ij}\right|}\dfrac{\partial\left|g_{ij}\right|}{\partial x^0}$ vanishes in (3-1). However, this leads to the following two conclusions:

① Because $\left|g_{ij}\right|$ is one of the six independent dynamical variables, this is obvious from the transformation (3-6), if $\dfrac{\partial |g_{ij}|}{\partial x^0}$ vanishes, then one of the six dynamical equations $\Theta^{ij}=0$ becomes an equation in which there is not any second time derivative term. Namely, the ten Einstein equations are divided into at least five equations of constraint in which there is not any second time derivative term and at most five equations of motion in which there are second time derivative terms.

② Because local Lorentz transformation (1-19) cannot leads to the change of $g_{ij}$, say nothing of $|g_{ij}|$, for the purpose that $\dfrac{\partial |g_{ij}|}{\partial x^0}$ vanishes, we therefore have to employ coordinate transformation $x^\mu = x^\mu(\tilde{x}^\nu)$, this means that the purpose can only be realized in some special coordinate system, namely, general relativity loses general covariance.

In spite of the above two conclusions, for the purpose that $\dfrac{\partial |g_{ij}|}{\partial x^0}$ vanishes, we simply choose

$$|g_{ij}| = 1. \qquad (3\text{-}9)$$

Although the condition (3-9) leads to a reduction of one in the six dynamical variables, it is allowable as long as there is a coordinate transformation such that (3-9) holds for arbitrary coordinate system. As an example, in the theory of the Yang-Mills field, although $A_3^a$ is a real dynamical variable, we can still choose so called space-axial gauge $A_3^a = 0$ [11]. The reason that space-axial gauge holds is that there exists a gauge transformation such that $A_3^a = 0$ holds for arbitrary gauge field $A_\mu^a$.

However, the condition (3-9) cannot guarantee that the quadratic term of time derivative in $L_{G0}$ is surely positive definitive. Because, although duo to (3-9), we have $e_{\hat{a}}^i e_{i,\lambda}^{\hat{a}} = \dfrac{1}{|^3 e|} \dfrac{\partial |^3 e|}{\partial x^\lambda} = \dfrac{1}{2} \dfrac{1}{|g_{ij}|} \dfrac{\partial |g_{ij}|}{\partial x^\lambda} = 0$, and the first term of (3-1) becomes

$$-\dfrac{2}{3}\left[e_0^\lambda e_{\hat{a}}^i \left(e_{i,\lambda}^{\hat{a}} - e_{\lambda,i}^{\hat{a}}\right)\right]^2 = -\dfrac{2}{3}\left(e_0^\lambda e_{\hat{a}}^i e_{\lambda,i}^{\hat{a}}\right)^2 = -\dfrac{2}{3}\left(e_0^0\right)^2 \left(e_{\hat{a}}^i e_{0,i}^{\hat{a}} - e_0^{\hat{b}} e_{\hat{b}}^j e_{\hat{a}}^i e_{j,i}^{\hat{a}}\right)^2,$$

we see that there are $e_0^{\hat{a}}$ in the above expressions. On the other hand, because there is not any time derivative term $e_{0,0}^{\hat{0}}$ and $e_{0,0}^{\hat{a}}$ in (2-4) and (2-5), hence, if we regard (2-4) and (2-5) as four holonomic equations of constraint about $e_0^{\hat{0}}$ and $e_0^{\hat{a}}$, then we therefore should solve (2-4) and (2-5) to obtain $e_0^{\hat{0}}$ and $e_0^{\hat{a}}$ as functions of $e_i^{\hat{b}}$, $e_{i,j}^{\hat{b}}$, and especially, $e_{i,0}^{\hat{b}}$. And then, substituting $e_0^{\hat{0}} = e_0^{\hat{0}}(e_i^{\hat{b}}, e_{i,j}^{\hat{b}}, e_{i,0}^{\hat{b}})$ and $e_0^{\hat{a}} = e_0^{\hat{a}}(e_i^{\hat{b}}, e_{i,j}^{\hat{b}}, e_{i,0}^{\hat{b}})$ to (3-1) for eliminating redundant variables. We see that maybe $e_{i,0}^{\hat{b}}$ appear again in the first term of $L_{G0}$ duo to $e_0^{\hat{a}} = e_0^{\hat{a}}(e_i^{\hat{b}}, e_{i,j}^{\hat{b}}, e_{i,0}^{\hat{b}})$. This discussion shows that if we only use (3-9) then it cannot guarantee that the quadratic term of time derivative in $L_G$ is surely positive definitive.

On the other hand, the above discussion shows yet that if $e_0^{\hat{a}} = 0$, from (2-2) and (3-2) then we see that this condition can be realized by choosing coordinate condition

$$g_{0i} = 0, \qquad (3\text{-}10)$$

then the first term of (3-1) vanishes and the quadratic term of time derivative in $L_{G0}$ is positive definitive. In fact, using (3-9) and (3-10), Landau and Lifschitz proved that the quadratic term of time derivative in $L_g$ of (1-1) is positive definitive[10].

Under the conditions (3-9) and (3-10), we have $\sqrt{|g_{ij}|} = |^3e| = h_0 = 1$ and $e_0^{\hat{a}} = 0$, which are two holonomic equations of constraint about $|^3e|$ and $e_0^{\hat{a}}$, and can be substituted directly to $L_G$ given by (3-1), from (1-16) we therefore can obtain an action

$$S_{(1)} = \frac{c^3}{16\pi G} \int d^4 x e_0^{\hat{0}} L_G \Big|_{|^3e|=1, e_0^{\hat{a}}=0}. \qquad (3\text{-}11)$$

From the above discussion we know that there is not negative kinetic energy term in (3-11). However, to substitute $|^3e|=1$ to $L_G$ ask a method that separates $|^3e|$ as an independent variable from the nine dynamical variables $e_i^{\hat{a}}$, we shall give such a method in §4 of this paper.

### 3.2 A coordinate condition insuring positive definiteness of the kinetic energy term in $L_{G0}$

According to (2-2) and (3-2) we have

$$\left(\sqrt{|g_{lm}|}\frac{g^{0\lambda}}{g^{00}}\right)_{,\lambda} = \sqrt{|g_{lm}|}_{,0} + \sqrt{|g_{lm}|}_{,i}\frac{g^{0i}}{g^{00}} + \sqrt{|g_{lm}|}\left(\frac{g^{0i}}{g^{00}}\right)_{,i}$$

$$= |^3e|_{,0} + |^3e|_{,i}\frac{-e_{\hat{0}}^0 e_{\hat{0}}^i}{-e_{\hat{0}}^0 e_{\hat{0}}^0} + |^3e|\left(\frac{-e_{\hat{0}}^0 e_{\hat{0}}^i}{-e_{\hat{0}}^0 e_{\hat{0}}^0}\right)_{,i} = |^3e|_{,0} + e_0^{\hat{0}} e_{\hat{0}}^i |^3e|_{,i} + |^3e|\left(e_0^{\hat{0}}\right)^2 \left(e_{\hat{0}}^i e_{\hat{0}, i}^0 - e_{\hat{0}, i}^i e_{\hat{0}}^0\right)$$

$$= e_0^{\hat{0}} |^3e| \left(e_{\hat{0}}^0 \frac{|^3e|_{,0}}{|^3e|} + e_{\hat{0}}^i \frac{|^3e|_{,i}}{|^3e|} + e_{\hat{0}, i}^i - e_{\hat{0}}^i e_{\hat{0}}^0 e_{\hat{0}, i}^0 \right) = e_0^{\hat{0}} |^3e| \left[e_{\hat{0}}^\lambda e_{\hat{a}}^i e_{i,\lambda}^{\hat{a}} + e_{\hat{0}, i}^\lambda \left(\delta_\lambda^i - e_{\hat{0}}^i e_\lambda^{\hat{0}}\right)\right]$$

$$= e_0^{\hat{0}} |^3e| \left(e_{\hat{0}}^\lambda e_{\hat{a}}^i e_{i,\lambda}^{\hat{a}} + e_{\hat{0}, i}^\lambda e_{\hat{a}}^i e_\lambda^{\hat{a}}\right) = e_0^{\hat{0}} |^3e| e_{\hat{0}}^\lambda e_{\hat{a}}^i \left(e_{i,\lambda}^{\hat{a}} - e_{\lambda, i}^{\hat{a}}\right), \qquad (3\text{-}12)$$

another proof of (3-12) can be found in §5 of this paper.

According to (3-12), the negative kinetic energy term in $L_{G0}$ reads

$$L_{GNK} = -\frac{2}{3}\left[e_{\hat{0}}^\lambda e_{\hat{a}}^i \left(e_{i,\lambda}^{\hat{a}} - e_{\lambda, i}^{\hat{a}}\right)\right]^2 = \frac{2}{3g}\left[\left(\sqrt{|g_{lm}|}\frac{g^{0\lambda}}{g^{00}}\right)_{,\lambda}\right]^2$$

$$= -\frac{2}{3}\left\{\sqrt{-g^{00}}\left[\frac{1}{2}\frac{|g_{lm}|_{,0}}{|g_{lm}|} + \frac{1}{2}\frac{g^{0i}}{g^{00}}\frac{|g_{lm}|_{,i}}{|g_{lm}|} + \left(\frac{g^{0i}}{g^{00}}\right)_{,i}\right]\right\}^2 \qquad (3\text{-}13)$$

$$= \frac{2}{3}g^{00}\left[\frac{1}{2}g^{lm}g_{lm,0} + \frac{1}{2}\frac{g^{0i}}{g^{00}}g^{lm}g_{lm,i} + \left(\frac{g^{0i}}{g^{00}}\right)_{,i}\right]^2 ;$$

Taking advantage of (3-13), we can calculate conveniently the negative kinetic energy term of gravitation field for given metric tensor $g_{\mu\nu}$. We investigate two examples.

① For the Robertson-Walker metric indicated by the line element

$$ds^2 = -d(ct)^2 + R^2(t)\left[\frac{dr^2}{1-kr^2} + r^2(d\theta^2 + \sin^2\theta d\varphi^2)\right],$$

we have $g^{00} = -1$, $g^{0i} = 0$, $|g_{ij}| = R^6(t)\dfrac{r^4 \sin^2\theta}{1-kr^2}$, according to (3-13), the corresponding negative kinetic energy term of the Robertson-Walker metric in total space reads

$$-\frac{2}{3}\left[e_{\hat{0}}^\lambda e_{\hat{a}}^i\left(e_{i,\lambda}^{\hat{a}} - e_{\lambda,i}^{\hat{a}}\right)\right]^2 = -\frac{6}{c^2}\frac{1}{R^2(t)}\left(\frac{dR(t)}{dt}\right)^2.$$

② For the Schwarzschild metric indicated by the line element

$$ds^2 = -\left(1-\frac{r_s}{r}\right)(dx^0)^2 + \frac{1}{1-\frac{r_s}{r}}dr^2 + r^2(d\theta^2 + \sin^2\theta d\varphi^2),$$

where $r_s = \dfrac{2GM}{c^2}$. We see that $\dfrac{\partial |g_{ij}|}{\partial (ct)} = \dfrac{\partial}{\partial (ct)}\left(\dfrac{r^4 \sin^2\theta}{1-\frac{r_s}{r}}\right) = 0$ in the area of $r_s < r$, hence, according to (3-13), in the area of $r_s < r$ the negative kinetic energy term of the Schwarzschild metric vanishes.

But the above form of the Schwarzschild metric cannot be continued into the area of $0 < r < r_s$. For continuing it into the area of $0 < r < r_s$, one has used a method of coordinate transformation and obtained some metrics, e.g., the Lemaitre and the Kruskal metrics. However, for the Lemaitre and the Kruskal metrics of the Schwarzschild solution, using (3-13), we can verify easily that there are corresponding negative kinetic energy terms of gravitation field in total space, respectively. (We don't discuss these questions in detail here.)

On the other hand, from (3-13) we see that if we choose

$$\left(\sqrt{|g_{lm}|}\frac{g^{0\lambda}}{g^{00}}\right)_{,\lambda} = 0, \tag{3-14}$$

then

$$e_{\hat{0}}^\lambda e_{\hat{a}}^i\left(e_{i,\lambda}^{\hat{a}} - e_{\lambda,i}^{\hat{a}}\right) = 0, \tag{3-15}$$

the quadratic term of time derivative in $L_{G0}$ given by (3-1) is thus positive definitive.

Hence, for the purpose obtaining a positive definitive quadratic term of time derivative in $L_{G0}$ given by (3-1), we present two groups of coordinate conditions: one is (3-9) and (3-10), another is (3-14). Of course, one can try to choose other "gauge-fixing term" for this purpose.

Substituting (3-15) to (2-4), (2-5) and (2-8), we obtain *the Einstein equations with the characteristic (3-15) under the condition (2-1)*, whose concrete forms no longer be written down here.

It is important that we can prove that the action

$$S_{(2)} = \frac{c^3}{16\pi G}\int |{}^4e|d^4x\, L_{GPK}, \tag{3-16}$$

$$L_{GPK} = L_G - L_{GNK} = L_{G0} - L_{GNK} + 2e_{\hat{0}}^0 e_{0,k}^{\hat{0}}U^k + L_{GV} \tag{3-17}$$

can leads to *the Einstein equations with the characteristic (3-15) under the condition (2-1)*, where $L_G$ is given by (2-3); especially, in $L_{GPK}$, time derivative terms only appear in the term

$$L_{G0} - L_{GNK} = \frac{1}{2} M^{ij}_{\hat{a}\hat{b}} e^{\mu}_{\hat{0}} \left( e^{\hat{a}}_{i,\mu} - e^{\hat{a}}_{\mu,i} \right) e^{\nu}_{\hat{0}} \left( e^{\hat{b}}_{j,\nu} - e^{\hat{b}}_{\nu,j} \right) + \frac{2}{3} \left[ e^{\lambda}_{\hat{0}} e^{i}_{\hat{a}} \left( e^{\hat{a}}_{i,\lambda} - e^{\hat{a}}_{\lambda,i} \right) \right]^2$$

$$= e^{\mu}_{\hat{0}} e^{\nu}_{\hat{0}} \left( \overline{g}^{il} \overline{g}^{jm} - \overline{g}^{ij} \overline{g}^{lm} \right) \Gamma_{\mu ij} \Gamma_{\nu lm} - \frac{2}{3} g^{00} \left[ \frac{1}{2} g^{lm} g_{lm,0} + \frac{1}{2} \frac{g^{0i}}{g^{00}} g^{lm} g_{lm,i} + \left( \frac{g^{0i}}{g^{00}} \right)_{,i} \right]^2.$$

It is obvious that there is not any negative kinetic energy term in $L_{GPK}$, thus, based on (3-11) or (3-16), we can try to realize quantization of general relativity by various methods of quantization, e.g., the Dirac-Bargmann method for a strange Lagrangian system, or the method of path integral. In this paper, we only discuss simply the method of canonical quantization.

## 4    The Hamiltonian representation

At first, in spite of the non-positive definiteness of the quadratic term of time derivative in $L_{G0}$ provisionally, we discuss the Hamiltonian representation of (2-3). As a first step of the Hamiltonian representation, we need 3+1 dimensional decomposition of space-time manifold, this can be realized by using the ADM decomposition[12]:

$$ds^2 = -\left( N^2 - h_{ij} N^i N^j \right) \left( dx^0 \right)^2 + 2 N_i dx^i dx^0 + h_{ij} dx^i dx^j, \qquad (4-1)$$

where $N_i = h_{ij} N^j$. For taking advantage of the forms of the foregoing formulas, we still use $g_{ij}$ to denote $h_{ij}$. Under (2-1), both (2-2) and (3-2) hold in this case, and especially we have

$$e^{\hat{0}}_0 = N, \quad e^{\hat{a}}_0 = e^{\hat{a}i} N_i = e^{\hat{a}}_i N^i, \quad e^i_{\hat{0}} = -e^0_{\hat{0}} N^i, \quad h^{ij} = \overline{g}^{ij} = e^i_{\hat{a}} e^{\hat{a}j}. \qquad (4-2)$$

From (3-3) we have

$$\sqrt{-g} L_{G0} = e^{\hat{0}}_0 |^3 e| e^{\mu}_{\hat{0}} e^{\nu}_{\hat{0}} \left( \overline{g}^{il} \overline{g}^{jm} - \overline{g}^{ij} \overline{g}^{lm} \right) \Gamma_{\mu ij} \Gamma_{\nu lm}$$

$$= \frac{\sqrt{|g_{ij}|}}{N} \left( \overline{g}^{il} \overline{g}^{jm} - \overline{g}^{ij} \overline{g}^{lm} \right) \left( \Gamma_{0ij} \Gamma_{0lm} - 2 N^n \Gamma_{0ij} \Gamma_{nlm} + N^k N^n \Gamma_{kij} \Gamma_{nlm} \right). \qquad (4-3)$$

Because there may be $e^{\hat{a}}_{i,0}$ (and, further, $g_{ij,0}$) in a Lagrangian of matter, e.g., the Lagrangian $L_D$ given by (1-11) of the Dirac field, for the sake of simpleness, we ignore Lagrangian of matter, and the momenta conjugate to $g_{ij}$ are

$$\pi^{ij} = \frac{\partial \left( \sqrt{-g} L_G \right)}{\partial g_{ij,0}} = \frac{\partial \left( \sqrt{-g} L_{G0} \right)}{\partial g_{ij,0}} = -\frac{\sqrt{|g_{ij}|}}{N} \left( \overline{g}^{il} \overline{g}^{jm} - \overline{g}^{ij} \overline{g}^{lm} \right) \left( \Gamma_{0lm} - N^n \Gamma_{nlm} \right). \qquad (4-4)$$

Using the DeWitt metric[13]

$$G_{ijlm} = \frac{1}{2} \frac{1}{\sqrt{|g_{ij}|}} \left( g_{il} g_{jm} + g_{im} g_{jl} - g_{ij} g_{lm} \right), \qquad (4-5)$$

from (4-4) we obtain

$$\Gamma_{0ij} = -N G_{ijlm} \pi^{lm} + N^k \Gamma_{kij}, \qquad (4-6)$$

and, further,

$$\sqrt{-g}L_{G0} = NG_{ijlm}\pi^{ij}\pi^{lm}. \tag{4-7}$$

Notice

$$g_{ij,0} = -2\Gamma_{0ij}g_{0i,j} + g_{0j,i} = 2NG_{ijlm}\pi^{lm} - 2N^k\Gamma_{kij} + N_{i,j} + N_{j,i},$$

we have

$$\pi^{ij}g_{ij,0} - \sqrt{-g}L_G$$
$$= NG_{ijlm}\pi^{ij}\pi^{lm} - 2N_{,k}\sqrt{|g_{ij}|}U^k - N\sqrt{|g_{ij}|}L_{GV} + \pi^{ij}(N_{i,j} + N_{j,i}) - 2N_i\,\overline{g}^{ij}\Gamma_{jlm}\pi^{lm},$$

hence, up to a total derivative, under the condition (2-1), the action of the system reads

$$S_{EH} = \frac{c^3}{16\pi G}\int|^4e|d^4x\,L_G = \frac{c^3}{16\pi G}\int d^4x\left(\pi^{ij}g_{ij,0} - NH_{\text{Hamiltonian}} - N_i H^i_{\text{Diffeomorphism}}\right), \tag{4-8}$$

$$H_{\text{Hamiltonian}} = G_{ijlm}\pi^{ij}\pi^{lm} + 2\left(\sqrt{|g_{ij}|}U^k\right)_{,k} - \sqrt{|g_{ij}|}L_{GV}, \tag{4-9}$$

$$H^i_{\text{Diffeomorphism}} = -2\left(\pi^{ij}{}_{,j} + \overline{g}^{ij}\Gamma_{jlm}\pi^{lm}\right). \tag{4-10}$$

The above forms show clearly the Diffeomorphism and the Hamiltonian constraints. The expression (4-10) of the Diffeomorphism constraint is just the same as the usual form[14], as for the usual form of the Hamiltonian constraint[13]

$$H_{\text{uf-Hamiltonian}} = G_{ijlm}\pi^{ij}\pi^{lm} + \sqrt{|g_{ij}|}R^{(3)}, \tag{4-11}$$

comparing (4-9) with (4-11) we see that the "kinetic energy term" in the two expressions is the same: both are $G_{ijlm}\pi^{ij}\pi^{lm}$.

Inasmuch as the success of the Ashtekar theory[15], we can try to simulate Ashtekar's method to introduce some new variables for simplifying the equations (4-9) and (4-10). However, according to known results of the Ashtekar theory, we can forecast that if we make the thing like so, then maybe we shall encounter some problem, e.g., a problem similar to that of "real condition" in the Ashtekar theory.

Before we try to simplify the equations (4-9) and (4-10), a more basic problem is that the kinetic energy term $G_{ijlm}\pi^{ij}\pi^{lm}$ in (4-9) and (4-11) is non-positive definitive, this is a consequence of (4-7) and (3-3). Concretely, according to (4-6) and (3-3), the negative term in $G_{ijlm}\pi^{ij}\pi^{lm}$ is $-\frac{2}{3}N\sqrt{|g_{ij}|}\left(G_{ijlm}\overline{g}^{ij}\pi^{lm}\right)^2$.

This characteristic can be shown more clearly by the transformation (3-6). If we substitute (3-7) to (3-3) and define

$$\pi_u = \frac{\partial\left(\sqrt{-g}L_G\right)}{\partial h_{u,0}} = \frac{\partial\left(\sqrt{-g}L_{G0}\right)}{\partial h_{u,0}}, u = 0, 1, 2, 3, 4, 5, \tag{4-12}$$

then through some derivation similar to (4-4) ~ (4-10), we can obtain a form of the Hamiltonian constraint:

$$H_{\text{Hamiltonian}} = -\frac{3}{8}h_0\pi_0^2 + \frac{1}{2h_0}\left[\frac{3}{4}h_1^2\pi_1^2 + \frac{1}{4}h_2^2\pi_2^2 + \frac{1}{h_2^4}\pi_3^2 + \frac{h_2^2}{h_1^2}(\pi_4 - h_3\pi_5)^2 + \frac{1}{h_1^2 h_2^2}\pi_5^2\right] \quad (4\text{-}13)$$
$$+ 2(h_0 U^k)_{,k} - h_0 L_{\text{GV}}.$$

This form shows clearly that the negative kinetic energy term in $H_{\text{Hamiltonian}}$ is $-\frac{3}{8}h_0\pi_0^2$.

Now we employ the coordinate condition (3-9), namely, $h_0 = 1$, then from (4-12) we can only get five momenta conjugate $\pi_u$, $u = 1, 2, 3, 4, 5$, because there is not $h_0$ in $L_{\text{G0}}$, and (4-13) becomes

$$H_{\text{Hamiltonian}} = \frac{1}{2}\left[\frac{3}{4}h_1^2\pi_1^2 + \frac{1}{4}h_2^2\pi_2^2 + \frac{1}{h_2^4}\pi_3^2 + \frac{h_2^2}{h_1^2}(\pi_4 - h_3\pi_5)^2 + \frac{1}{h_1^2 h_2^2}\pi_5^2\right] + 2U^k_{,k} - L_{\text{GV}}, \quad (4\text{-}14)$$

we see that the kinetic energy term in the Hamiltonian constraint given by (4-14) is positive definitive.

As for the potential energy term $2U^k_{,k} - L_{\text{GV}}$ in (4-14), we have to add some gauge conditions such that $e_i^{\hat{a}}$ can be expressed by $g_{ij}$, and, further, by $h_u$ ($u$=0, 1, 2, 3, 4, 5) according to (3-7):

$$e_i^{\hat{a}} = e_i^{\hat{a}}(g_{lm}) = e_i^{\hat{a}}(h_u).$$

For this purpose, we can generalize (2-1) to the form

$$e_{\hat{\alpha}\mu} = 0, \quad \alpha < \mu, \quad (4\text{-}15)$$

combining the last formula in (3-2), $e_{\hat{a}i}$ is thus a triangular matrix:

$$\begin{bmatrix} e_{\hat{1}1} & e_{\hat{1}2} & e_{\hat{1}3} \\ e_{\hat{2}1} & e_{\hat{2}2} & e_{\hat{2}3} \\ e_{\hat{3}1} & e_{\hat{3}2} & e_{\hat{3}3} \end{bmatrix} = \begin{bmatrix} \dfrac{\sqrt{|g_{ij}|}}{\sqrt{g_{22}g_{33} - (g_{23})^2}} & 0 & 0 \\ -\dfrac{g_{23}g_{31} - g_{12}g_{33}}{\sqrt{g_{33}}\sqrt{g_{22}g_{33} - (g_{23})^2}} & \dfrac{\sqrt{g_{22}g_{33} - (g_{23})^2}}{\sqrt{g_{33}}} & 0 \\ \dfrac{g_{31}}{\sqrt{g_{33}}} & \dfrac{g_{23}}{\sqrt{g_{33}}} & \sqrt{g_{33}} \end{bmatrix}. \quad (4\text{-}16)$$

The conclusion that the above form of $e_{\hat{a}i}$ always exists has been proved in many literatures. In fact, according to the last formula in (3-2), (3-6) thus hold, then the form of $e_{\hat{a}i}$ given by (4-16) is so called the Cholesky decomposition in algebra.

And, further, according to (3-7) we have

$$\begin{bmatrix} e_{\hat{1}1} & e_{\hat{1}2} & e_{\hat{1}3} \\ e_{\hat{2}1} & e_{\hat{2}2} & e_{\hat{2}3} \\ e_{\hat{3}1} & e_{\hat{3}2} & e_{\hat{3}3} \end{bmatrix} = \begin{bmatrix} \sqrt[3]{h_0 h_1}\,\dfrac{1}{h_1} & 0 & 0 \\ -\sqrt[3]{h_0 h_1}\,\dfrac{h_4}{h_2} & \sqrt[3]{h_0 h_1}\,\dfrac{1}{h_2} & 0 \\ -\sqrt[3]{h_0 h_1}\,h_2(h_3 h_4 + h_5) & \sqrt[3]{h_0 h_1}\,h_2 h_3 & \sqrt[3]{h_0 h_1}\,h_2 \end{bmatrix}. \quad (4\text{-}17)$$

Based on (4-16), from $g_{0i} = e_0^{\hat{a}} e_{\hat{a}i}$ in (3-2) we have

$$e_{\hat{1}0} = \frac{g_{01}[g_{22}g_{33} - (g_{23})^2] + g_{02}(g_{23}g_{31} - g_{12}g_{33}) + g_{03}(g_{12}g_{23} - g_{22}g_{31})}{\sqrt{|g_{ij}|}\sqrt{g_{22}g_{33} - (g_{23})^2}},$$

$$e_{\hat{2}0} = \frac{g_{02}g_{33} - g_{03}g_{23}}{\sqrt{g_{33}}\sqrt{g_{22}g_{33} - (g_{23})^2}}, \quad e_{\hat{3}0} = \frac{g_{03}}{\sqrt{g_{33}}}.$$

(4-18)

Using (2-1) and (2-2) we have

$$e_{\hat{0}}^0 = \left(e_0^{\hat{0}}\right)^{-1} = \sqrt{-g^{00}}, \quad e_{\hat{0}}^i = \frac{-g^{0i}}{\sqrt{-g^{00}}}.$$

(4-19)

Now we can substitute (4-17) to the potential energy term $2U^k{}_{,k} - L_{GV}$ in (4-14), and consider $h_0 = 1$, the last result of $2U^k{}_{,k} - L_{GV}$ obtained by computer is in the Appendix of this paper.

As well-known, after realizing canonical quantization, $\pi^{ij}(x) \to \frac{\hbar}{i}\frac{\delta}{\delta g_{ij}(x)}$ in (4-11) and (4-11) becomes an equation of constraint for wave function $\Psi[g_{ij}(x)]$:

$$H_{\text{uf-Hamiltonian}}\Psi[g_{ij}(x)] = \left(G_{ijlm}\frac{\hbar}{i}\frac{\delta}{\delta g_{ij}(x)}\frac{\hbar}{i}\frac{\delta}{\delta g_{lm}(x)} + \sqrt{|g_{ij}|}R^{(3)}\right)\Psi[g_{ij}(x)] = 0,$$

this is so called the Wheeler-DeWitt equation. Similarly, after realizing canonical quantization, (4-14) becomes an equation of constraint for wave function $\Psi[h_u(x)]$:

$$\left(\frac{1}{2}\left(\frac{3}{4}(h_1^2\pi_1^2)_W + \frac{1}{4}(h_2^2\pi_2^2)_W + \frac{1}{h_2^4}\pi_3^2 + \frac{h_2^2}{h_1^2}(\pi_4 - h_3\pi_5)^2 + \frac{1}{h_1^2 h_2^2}\pi_5^2\right) + 2U^k{}_{,k} - L_{GV}\right)\Psi[h_u(x)] = 0,$$

(4-20)

in which $\pi_u(x) \to \frac{\hbar}{i}\frac{\delta}{\delta h_u(x)}$ ($u$=1, 2, 3, 4, 5); $(h_i^2\pi_i^2)_W$ ($i = 1, 2$) means the Weyl ordering:

$$(h_i^2\pi_i^2)_W = \frac{1}{6}\left(h_i^2\pi_i^2 + h_i\pi_i^2 h_i + \pi_i h_i^2 \pi_i + \pi_i^2 h_i^2 + h_i\pi_i h_i \pi_i + \pi_i h_i \pi_i h_i\right) \quad (i = 1, 2).$$

Because (4-17) provides a method that separates $|^3e| = h_0$ as an independent variable from the nine dynamical variables $e_i^{\hat{a}}$ (in which there are only six independent variables), we now can discuss the theory of canonical quantization of general relativity based on the action (3-11).

If we use the action (3-11) to realize canonical quantization of general relativity, then the Hamiltonian constraint is still given by (4-20). But because there is not $e_0^{\hat{a}}$ in (3-11), we cannot obtain directly the Diffeomorphism constraint. On the other hand, after realizing canonical quantization and obtaining the Hamiltonian $H_{(1)}$ from the action (3-11), all the Diffeomorphism constraint $H^i_{\text{Diffeomorphism}}$ in which all the variables become the corresponding operators are commutative with $H_{(1)}$:

$$\left[H^i_{\text{Diffeomorphism}}, H_{(1)}\right] = 0.$$

Therefore, all $H^i_{\text{Diffeomorphism}}$ are conservation quantities of the theory and can be diagonalized with $H_{(1)}$ at the same time. Hence, we can pick out such $\Psi[h_u(x)]$ that satisfies $H^i_{\text{Diffeomorphism}}\Psi[h_u(x)] = 0$ as physically significant wave function. By this process, we obtain the Diffeomorphism constraint again. This method is the same as that of processing the Gaussian law under the temporal gauge in QED[16].

Similar to the case of $e_0^{\hat{a}}$, we cannot obtain directly an equation of constraint corresponding to $h_0$ from the action (3-11), because there is not $h_0$ in (3-11) as well. However, we can obtain it by the method similar to that of obtaining the Diffeomorphism constraint, which is discussed in the above paragraph.

We no longer write down the commutation relations and the equations of motion of the operators in the theory of canonical quantization obtained by the action (3-11) here.

We now discuss the theory of canonical quantization of general relativity based on the action (3-16).

For the action (3-16), substituting (4-17) to (3-17) and considering (4-2), we have

$$S_{(2)} = \frac{c^3}{16\pi G}\int d^4x\, Nh_0 L_{\text{GPK}}(h_u; h_{u,\lambda}; N, N_i, h_0; N_{,i}, N_{i,j}, h_{0,i}), \quad u = 1,2,3,4,5, \quad (4\text{-}21)$$

defining

$$\pi_u = \frac{\partial(Nh_0 L_{\text{GPK}})}{\partial h_{u,0}} = \frac{\partial(Nh_0(L_{\text{G0}} - L_{\text{GNK}}))}{\partial h_{u,0}}, \quad u = 1,2,3,4,5, \quad (4\text{-}22)$$

and from (4-22) we obtain $h_{u,0}$ as the functions of $\pi_v$:

$$h_{u,0} = h_{u,0}(\pi_v), \quad u, v = 1,2,3,4,5. \quad (4\text{-}23)$$

Substituting (4-23) to (4-21), we have

$$S = \frac{c^3}{16\pi G}\int d^4x\, Nh_0 L_{\text{GPK}}(\pi_u; h_v; h_{v,i}; N, N_i, h_0; N_{,i}, N_{i,j}, h_{0,i}), \quad u, v = 1,2,3,4,5. \quad (4\text{-}24)$$

From the above expression we can obtain five constraints:

The Hamiltonian constraint: $\dfrac{\partial(Nh_0 L_{\text{GPK}})}{\partial N} - \partial_i \dfrac{\partial(Nh_0 L_{\text{GPK}})}{\partial N_{,i}} = 0$, (4-25)

The Diffeomorphism constraint: $\dfrac{\partial(Nh_0 L_{\text{GPK}})}{\partial N_{,i}} - \partial_j \dfrac{\partial(Nh_0 L_{\text{GPK}})}{\partial N_{i,j}} = 0$, (4-26)

The $h_0$ constraint: $\dfrac{\partial(Nh_0 L_{\text{GPK}})}{\partial h_0} - \partial_i \dfrac{\partial(Nh_0 L_{\text{GPK}})}{\partial h_{0,i}} = 0$. (4-27)

After realizing canonical quantization, the commutation relations are

$$[h_u(t,\mathbf{x}), \pi_v(t,\mathbf{x}')] = i\hbar\delta_{uv}\delta^3(\mathbf{x}-\mathbf{x}'); \quad [h_u(t,\mathbf{x}), h_v(t,\mathbf{x}')] = 0;$$
$$[\pi_u(t,\mathbf{x}), \pi_v(t,\mathbf{x}')] = 0; \quad u, v = 1,2,3,4,5. \quad (4\text{-}28)$$

(4-25) ~ (4-27) become five equations of constraint for wave function $\Psi[h_u(x)]$, the five equations of motion of the operators read

$$i\hbar\dot{\pi}_u(t,\mathbf{x}) = [\pi_u(t,\mathbf{x}), H_{(2)}]; \quad u = 1,2,3,4,5. \quad (4\text{-}29)$$

In (4-29),

$$H_{(2)} = \int d^4x \left( \sum_{u=1}^{5} \pi_u h_{u,0}(\pi_v) - Nh_0 L_{\text{GPK}}(\pi_u; h_v; h_{v,i}; N, N_i, h_0; N_{,i}, N_{i,j}, h_{0,i}) \right). \quad (4\text{-}30)$$

All the concrete forms of (4-21) ~ (4-30) obtained by computer are complicated.

We therefore present two theories of canonical quantization of general relativity under two different groups of coordinate conditions: one is (3-9) and (3-10), another is (3-14), respectively. The common characteristics of the two theories are that the kinetic energy terms in both the actions and both the Hamiltonian constraints are positive definitive.

## 5 A group of gauge conditions making there is not any second time derivative term in the ten Einstein equations

We investigate two groups of tetrads: $\{\tilde{e}_\mu^{\hat{\alpha}}\}$ and $\{e_\mu^{\hat{\alpha}}\}$, for which the Schwinger time gauge condition holds, namely, (2-1) holds for $\{e_\mu^{\hat{\alpha}}\}$ and

$$\tilde{e}_{\hat{a}}^0 = 0, \quad a = 1, 2, 3. \quad (5\text{-}1)$$

Hence, a local Lorentz transformation $\Lambda_{\hat{\beta}}^{\hat{\alpha}}$ between such $\{\tilde{e}_\mu^{\hat{\alpha}}\}$ and $\{e_\mu^{\hat{\alpha}}\}$ has the characteristics:

$$\Lambda_{\hat{0}}^{\hat{0}}(x) = 1, \quad \Lambda_{\hat{0}}^{\hat{a}}(x) = 0, \quad \Lambda_{\hat{a}}^{\hat{0}}(x) = 0, \quad \eta^{\hat{c}\hat{d}} \Lambda_{\hat{c}}^{\hat{a}} \Lambda_{\hat{d}}^{\hat{b}} = \eta^{\hat{a}\hat{b}}; \quad (5\text{-}2)$$

Under the above special local Lorentz transformation, the relation between $\{\tilde{e}_\mu^{\hat{\alpha}}\}$ and $\{e_\mu^{\hat{\alpha}}\}$ reads:

$$e_{\hat{0}}^\mu = \tilde{e}_{\hat{0}}^\mu, \quad e_\mu^{\hat{0}} = \tilde{e}_\mu^{\hat{0}}, \quad e_\mu^{\hat{a}} = \Lambda_{\hat{b}}^{\hat{a}}(x) \tilde{e}_\mu^{\hat{b}}. \quad (5\text{-}3)$$

According to $e^{\hat{b}i} e_i^{\hat{c}} = e^{\hat{b}\lambda} e_\lambda^{\hat{c}} = \eta^{\hat{b}\hat{c}}$ we have

$$\tilde{e}^{\hat{a}i} \tilde{e}_{i,\lambda}^{\hat{a}} = \overline{\Lambda}_{\hat{b}}^{\hat{a}} \overline{\Lambda}_{\hat{c}}^{\hat{a}} e^{\hat{b}i} e_{i,\lambda}^{\hat{c}} + \overline{\Lambda}_{\hat{b}}^{\hat{a}} \overline{\Lambda}_{\hat{c},\lambda}^{\hat{a}} e^{\hat{b}i} e_i^{\hat{c}} = \overline{\Lambda}_{\hat{b}}^{\hat{a}} \overline{\Lambda}_{\hat{c}}^{\hat{a}} e^{\hat{b}i} e_{i,\lambda}^{\hat{c}} + \overline{\Lambda}_{\hat{b}}^{\hat{a}} \overline{\Lambda}_{\hat{c},\lambda}^{\hat{a}} \eta^{\hat{b}\hat{c}}, \quad (5\text{-}4)$$

and using $\overline{\Lambda}_{\hat{c}}^{\hat{a}} \eta^{\hat{b}\hat{c}} = \Lambda_{\hat{c}}^{\hat{b}} \eta^{\hat{a}\hat{c}}$, which can be proved by (5-2), $\overline{\Lambda}_{\hat{b}}^{\hat{a}} \overline{\Lambda}_{\hat{c},\lambda}^{\hat{a}} \eta^{\hat{b}\hat{c}} = \overline{\Lambda}_{\hat{b}}^{\hat{a}} \Lambda_{\hat{c},\lambda}^{\hat{b}} \eta^{\hat{a}\hat{c}} = \Lambda_{\hat{b}}^{\hat{c}} \overline{\Lambda}_{\hat{c},\lambda}^{\hat{a}} \eta^{\hat{b}\hat{a}}$,

we therefore have

$$\overline{\Lambda}_{\hat{b}}^{\hat{a}} \overline{\Lambda}_{\hat{c},\lambda}^{\hat{a}} \eta^{\hat{b}\hat{c}} = \frac{1}{2}\left( \overline{\Lambda}_{\hat{b}}^{\hat{a}} \Lambda_{\hat{c},\lambda}^{\hat{b}} + \Lambda_{\hat{c}}^{\hat{b}} \overline{\Lambda}_{\hat{b},\lambda}^{\hat{a}} \right) \eta^{\hat{a}\hat{c}} = \frac{1}{2} \left( \overline{\Lambda}_{\hat{b}}^{\hat{a}} \Lambda_{\hat{c}}^{\hat{b}} \right)_{,\lambda} \eta^{\hat{a}\hat{c}} = \frac{1}{2} \delta_{\hat{c},\lambda}^{\hat{a}} \eta^{\hat{a}\hat{c}} = 0,$$

(5-4) thus becomes

$$\tilde{e}^{\hat{a}i} \tilde{e}_{i,\lambda}^{\hat{a}} = \overline{\Lambda}_{\hat{b}}^{\hat{a}} \overline{\Lambda}_{\hat{c}}^{\hat{a}} e^{\hat{b}i} e_{i,\lambda}^{\hat{c}} \equiv \Omega_\lambda^{(a)}, \quad (5\text{-}5)$$

and using (5-3) we have

$$\tilde{e}_{\hat{0}}^\lambda \tilde{e}^{\hat{a}i} \tilde{e}_{i,\lambda}^{\hat{a}} = e_{\hat{0}}^\lambda \overline{\Lambda}_{\hat{b}}^{\hat{a}} \overline{\Lambda}_{\hat{c}}^{\hat{a}} e^{\hat{b}i} e_{i,\lambda}^{\hat{c}} = e_{\hat{0}}^\lambda \Omega_\lambda^{(a)}. \quad (5\text{-}6)$$

According to (5-3) and $e_{\hat{0}}^\lambda e_\lambda^{\hat{c}} = \delta_{\hat{0}}^{\hat{c}} = 0$ we have

$$\tilde{e}_{\hat{0}}^\lambda \tilde{e}^{\hat{a}i} \tilde{e}_{\lambda,i}^{\hat{a}} = e_{\hat{0}}^\lambda \left( \overline{\Lambda}_{\hat{b}}^{\hat{a}} \overline{\Lambda}_{\hat{c}}^{\hat{a}} e^{\hat{b}i} e_{\lambda,i}^{\hat{c}} + \overline{\Lambda}_{\hat{b}}^{\hat{a}} \overline{\Lambda}_{\hat{c},i}^{\hat{a}} e^{\hat{b}i} e_\lambda^{\hat{c}} \right) = e_{\hat{0}}^\lambda \overline{\Lambda}_{\hat{b}}^{\hat{a}} \overline{\Lambda}_{\hat{c}}^{\hat{a}} e^{\hat{b}i} e_{\lambda,i}^{\hat{c}} \equiv \omega^{(a)}. \quad (5\text{-}7)$$

Especially,

$$\sum_{a=1}^{3}\left(e_{\hat{0}}^{\lambda}\Omega_{\lambda}^{(a)}-\omega^{(a)}\right)=\widetilde{e}_{\hat{0}}^{\lambda}\widetilde{e}_{\hat{a}}^{i}\left(\widetilde{e}_{i,\lambda}^{\hat{a}}-\widetilde{e}_{\lambda,i}^{\hat{a}}\right)=\sum_{a=1}^{3}\overline{\Lambda}_{\hat{b}}^{\hat{a}}\overline{\Lambda}_{\hat{c}}^{\hat{a}}e_{\hat{0}}^{\lambda}\left(e^{\hat{b}i}e_{i,\lambda}^{\hat{c}}-e^{\hat{b}i}e_{\lambda,i}^{\hat{c}}\right)$$
$$=\eta_{\hat{a}\hat{d}}\overline{\Lambda}_{\hat{b}}^{\hat{a}}\overline{\Lambda}_{\hat{c}}^{\hat{d}}e_{\hat{0}}^{\lambda}\left(e^{\hat{b}i}e_{i,\lambda}^{\hat{c}}-e^{\hat{b}i}e_{\lambda,i}^{\hat{c}}\right)=\eta_{\hat{b}\hat{c}}e_{\hat{0}}^{\lambda}\left(e^{\hat{b}i}e_{i,\lambda}^{\hat{c}}-e^{\hat{b}i}e_{\lambda,i}^{\hat{c}}\right)=e_{\hat{0}}^{\lambda}e_{\hat{a}}^{i}\left(e_{i,\lambda}^{\hat{a}}-e_{\lambda,i}^{\hat{a}}\right),$$
(5-8)

this means that $e_{\hat{0}}^{\lambda}e_{\hat{a}}^{i}\left(e_{i,\lambda}^{\hat{a}}-e_{\lambda,i}^{\hat{a}}\right)$ is an invariable under the transformation (5-2).

We now designate that the tetrad $\{e_{\mu}^{\hat{\alpha}}\}$ in (5-3) satisfy (4-15). According to $e_{i}^{\hat{0}}=0$ in (2-2), (4-16), (4-18) and (4-19) we see that $\{e_{\mu}^{\hat{\alpha}}\}$ have been expressed as functions of metric tensor $g_{\mu\nu}$:

$$e_{\mu}^{\hat{\alpha}}=e_{\mu}^{\hat{\alpha}}(g_{\rho\sigma});\qquad(5\text{-}9)$$

According to (5-8), (2-2), (4-16), (4-18) and (4-19) we have

$$\widetilde{e}_{\hat{0}}^{\lambda}\widetilde{e}_{\hat{a}}^{i}\left(\widetilde{e}_{i,\lambda}^{\hat{a}}-\widetilde{e}_{\lambda,i}^{\hat{a}}\right)=e_{\hat{0}}^{\lambda}e_{\hat{a}}^{i}\left(e_{i,\lambda}^{\hat{a}}-e_{\lambda,i}^{\hat{a}}\right)=e_{\hat{0}}^{0}\left[\frac{\left|{}^{3}e\right|_{,0}}{\left|{}^{3}e\right|}+e_{\hat{0}}^{\hat{0}}e_{0}^{i}\frac{\left|{}^{3}e\right|_{,i}}{\left|{}^{3}e\right|}+\left(e_{0}^{\hat{0}}e_{\hat{0}}^{i}\right)_{,i}\right]$$
$$=\sqrt{-g^{00}}\left[\frac{1}{2}\frac{\left|g_{lm}\right|_{,0}}{\left|g_{lm}\right|}+\frac{1}{2}\frac{g^{0i}}{g^{00}}\frac{\left|g_{lm}\right|_{,i}}{\left|g_{lm}\right|}+\left(\frac{g^{0i}}{g^{00}}\right)_{,i}\right]=\frac{1}{\sqrt{-g}}\left(\sqrt{\left|g_{lm}\right|}\frac{g^{0\lambda}}{g^{00}}\right)_{,\lambda},$$

(3-12) thus be proved again for the transformation (5-2).

Now that $\{\widetilde{e}_{\mu}^{\hat{\alpha}}\}$ satisfy (5-1), the characteristics (2-2) thus holds yet for $\{\widetilde{e}_{\mu}^{\hat{\alpha}}\}$. And, further, we have

$$\widetilde{e}_{\hat{0}}^{\mu}=\frac{-g^{0\mu}}{\sqrt{-g^{00}}},\quad\widetilde{e}_{\mu}^{\hat{0}}=\frac{\delta_{\mu}^{0}}{\sqrt{-g^{00}}}.\qquad(5\text{-}10)$$

(5-10) shows that $\widetilde{e}_{\mu}^{\hat{0}}$ as one of the four components of $\widetilde{e}_{\mu}^{\hat{\alpha}}$ has been completely determined for the metric tensor $g_{\mu\nu}$. On the other hand, in Ref. [17], it is pointed out that, in this case (i.e., one of the four vectors $\widetilde{e}_{\mu}^{\hat{\alpha}}$ has been completely determined for the metric tensor $g_{\mu\nu}$), one can choose other $\widetilde{e}_{\mu}^{\hat{k}}$, $k=1,2,3$ such that the Ricci's coefficients of rotation $\widetilde{r}_{\hat{\alpha}\hat{\beta}\hat{\gamma}}$ satisfy

$$\widetilde{r}_{\hat{0}\hat{a}\hat{b}}+\widetilde{r}_{\hat{0}\hat{b}\hat{a}}=0,\ a\neq b;\ a,b=1,2,3,$$

concretely,

$$\widetilde{r}_{\hat{0}\hat{1}\hat{2}}+\widetilde{r}_{\hat{0}\hat{2}\hat{1}}=0,\ \widetilde{r}_{\hat{0}\hat{1}\hat{3}}+\widetilde{r}_{\hat{0}\hat{3}\hat{1}}=0,\ \widetilde{r}_{\hat{0}\hat{2}\hat{3}}+\widetilde{r}_{\hat{0}\hat{3}\hat{2}}=0.\qquad(5\text{-}11)$$

In Ref. [17], (5-11) is called *the simplest case*, because it leads to that the independent components of the Ricci's coefficients of rotation reduce to 21 from at most 24 (see (1-8)) in the 4-dimensional Riemannian geometry, we therefore call (5-11) *the simplest constraint conditions*.

In fact, from the definition (1-4) of the Ricci's coefficients of rotation and (5-1) we have

$$\widetilde{r}_{\hat{0}\hat{a}\hat{b}} + \widetilde{r}_{\hat{0}\hat{b}\hat{a}} = \widetilde{e}_0^{\lambda} \left[ \widetilde{e}_{\hat{a}}^{i} \left( \widetilde{e}_{\hat{b}i,\lambda} - \widetilde{e}_{\hat{b}\lambda,i} \right) + \widetilde{e}_{\hat{b}}^{i} \left( \widetilde{e}_{\hat{a}i,\lambda} - \widetilde{e}_{\hat{a}\lambda,i} \right) \right]. \tag{5-12}$$

If there is a group of tetrads $\{e_\mu^{\hat{\alpha}}\}$ has the characteristic $e_{\hat{a}}^0 = 0$, $a = 1,2,3$ as well but $r_{\hat{0}\hat{a}\hat{b}} + r_{\hat{0}\hat{b}\hat{a}} \neq 0$ ($a \neq b$; $a, b = 1,2,3$), for example, for the tetrad $\{e_\mu^{\hat{\alpha}}\}$ satisfying (4-15), we have $r_{\hat{0}\hat{a}\hat{b}} + r_{\hat{0}\hat{b}\hat{a}} \neq 0$ ($a \neq b$; $a, b = 1,2,3$), then under the transformation (5-2), we have

$$\begin{aligned} r_{\hat{0}\hat{a}\hat{b}} + r_{\hat{0}\hat{b}\hat{a}} &= e_0^{\lambda} \left[ e_{\hat{a}}^{i} \left( e_{\hat{b}i,\lambda} - e_{\hat{b}\lambda,i} \right) + e_{\hat{b}}^{i} \left( e_{\hat{a}i,\lambda} - e_{\hat{a}\lambda,i} \right) \right] \\ &= \overline{\Lambda}_{\hat{a}}^{\hat{c}} \overline{\Lambda}_{\hat{b},\lambda}^{\hat{d}} \widetilde{e}_0^{\lambda} \widetilde{e}_{\hat{c}}^{i} \widetilde{e}_{\hat{d}i} - \overline{\Lambda}_{\hat{a}}^{\hat{c}} \overline{\Lambda}_{\hat{b},i}^{\hat{d}} \widetilde{e}_0^{\lambda} \widetilde{e}_{\hat{c}}^{i} \widetilde{e}_{\hat{d}\lambda} + \overline{\Lambda}_{\hat{a},\lambda}^{\hat{c}} \overline{\Lambda}_{\hat{b}}^{\hat{d}} \widetilde{e}_0^{\lambda} \widetilde{e}_{\hat{d}}^{i} \widetilde{e}_{\hat{c}i} - \overline{\Lambda}_{\hat{a},i}^{\hat{c}} \overline{\Lambda}_{\hat{b}}^{\hat{d}} \widetilde{e}_0^{\lambda} \widetilde{e}_{\hat{d}}^{i} \widetilde{e}_{\hat{c}\lambda} \\ &\quad + \overline{\Lambda}_{\hat{a}}^{\hat{c}} \overline{\Lambda}_{\hat{b}}^{\hat{d}} \widetilde{e}_0^{\lambda} \left[ \widetilde{e}_{\hat{c}}^{i} \left( \widetilde{e}_{\hat{d}i,\lambda} - \widetilde{e}_{\hat{d}\lambda,i} \right) + \widetilde{e}_{\hat{d}}^{i} \left( \widetilde{e}_{\hat{c}i,\lambda} - \widetilde{e}_{\hat{c}\lambda,i} \right) \right] \\ &= \overline{\Lambda}_{\hat{a}}^{\hat{c}} \overline{\Lambda}_{\hat{b}}^{\hat{d}} \left( \widetilde{r}_{\hat{0}\hat{c}\hat{d}} + \widetilde{r}_{\hat{0}\hat{d}\hat{c}} \right), \end{aligned}$$

where the formulas $\overline{\Lambda}_{\hat{a}}^{\hat{c}} \Lambda_{\hat{c}}^{\hat{b}} = \delta_{\hat{a}}^{\hat{b}}$, $\widetilde{e}_{\hat{c}}^{i} \widetilde{e}_{\hat{d}i} = \eta_{\hat{c}\hat{d}}$, $\eta_{\hat{c}\hat{d}} \overline{\Lambda}_{\hat{a}}^{\hat{c}} \overline{\Lambda}_{\hat{b},\lambda}^{\hat{d}} + \eta_{\hat{c}\hat{d}} \overline{\Lambda}_{\hat{a},\lambda}^{\hat{c}} \overline{\Lambda}_{\hat{b}}^{\hat{d}} = \left( \eta_{\hat{c}\hat{d}} \overline{\Lambda}_{\hat{a}}^{\hat{c}} \overline{\Lambda}_{\hat{b}}^{\hat{d}} \right)_{,\lambda} = (\eta_{\hat{a}\hat{b}})_{,\lambda} = 0$ and $\widetilde{e}_0^{\lambda} \widetilde{e}_{\hat{d}\lambda} = \eta_{\hat{0}\hat{d}} = 0$ are used. From the above formula we have

$$\widetilde{r}_{\hat{0}\hat{a}\hat{b}} + \widetilde{r}_{\hat{0}\hat{b}\hat{a}} = \Lambda_{\hat{a}}^{\hat{c}} \Lambda_{\hat{b}}^{\hat{d}} \left( r_{\hat{0}\hat{c}\hat{d}} + r_{\hat{0}\hat{d}\hat{c}} \right); \tag{5-13}$$

We see that if the non diagonal elements of the $3 \times 3$ symmetric matrix $\left[ r_{\hat{0}\hat{c}\hat{d}} + r_{\hat{0}\hat{d}\hat{c}} \right]$ are not zero, then we can make an orthogonal transformation by an orthogonal matrix $\Lambda_{\hat{a}}^{\hat{b}}$ such that the new 3 $\times$ 3 symmetric matrix $\left[ \widetilde{r}_{\hat{0}\hat{c}\hat{d}} + \widetilde{r}_{\hat{0}\hat{d}\hat{c}} \right]$ is diagonal, namely, $\left[ \widetilde{r}_{\hat{0}\hat{c}\hat{d}} + \widetilde{r}_{\hat{0}\hat{d}\hat{c}} \right]$ satisfy (5-11). Concretely, from (5-13) we see that $\left[ \widetilde{r}_{\hat{0}\hat{c}\hat{d}} + \widetilde{r}_{\hat{0}\hat{d}\hat{c}} \right]$ becomes a diagonal matrix if and only if

$$\Lambda_{\hat{1}}^{\hat{c}} \Lambda_{\hat{2}}^{\hat{d}} \left( r_{\hat{0}\hat{c}\hat{d}} + r_{\hat{0}\hat{d}\hat{c}} \right) = 0, \quad \Lambda_{\hat{1}}^{\hat{c}} \Lambda_{\hat{3}}^{\hat{d}} \left( r_{\hat{0}\hat{c}\hat{d}} + r_{\hat{0}\hat{d}\hat{c}} \right) = 0, \quad \Lambda_{\hat{2}}^{\hat{c}} \Lambda_{\hat{3}}^{\hat{d}} \left( r_{\hat{0}\hat{c}\hat{d}} + r_{\hat{0}\hat{d}\hat{c}} \right) = 0. \tag{5-14}$$

If (5-11) hold, then according to (5-12), $\widetilde{S}^{\hat{0}i\hat{a}}$ given by (2-6) but in which $\{\widetilde{e}_\mu^{\hat{\alpha}}\}$ is now as basic variables becomes

$$\widetilde{S}^{\hat{0}i\hat{a}} = \widetilde{e}^{\hat{a}i} \widetilde{e}_0^{\lambda} \sum_{\substack{b=1 \\ b \neq a}}^{3} \widetilde{e}^{\hat{b}j} \left( \widetilde{e}_{j,\lambda}^{\hat{b}} - \widetilde{e}_{\lambda,j}^{\hat{b}} \right), \tag{5-15}$$

and according to (1-27) we can prove that there is not any second time derivative term in the three equations $\widetilde{\Theta}^{ij} = 0$ ($i \neq j$) in which $\{\widetilde{e}_\mu^{\hat{\alpha}}\}$ is as basic variables, this means that all second time derivative terms are eliminated in the three of the six equations of motion $\widetilde{\Theta}^{ij} = 0$.

Now that we have designated that $\{e_\mu^{\hat{\alpha}}\}$ satisfy (4-15), of course, we have $r_{\hat{0}\hat{a}\hat{b}} + r_{\hat{0}\hat{b}\hat{a}} \neq 0$ ($a \neq b$; $a, b = 1,2,3$). On the other hand, we now ask that $\{\widetilde{e}_\mu^{\hat{\alpha}}\}$ satisfy (5-11), according to (5-2), (5-13) and (5-14), a special local Lorentz transformation $\Lambda^{\hat{\alpha}}_{\hat{\beta}}$ between $\{e^{\hat{\alpha}}_{\mu}\}$ and $\{\tilde{e}^{\hat{\alpha}}_{\mu}\}$ is fully determined.

(5-9) shows that $\{e^{\hat{\alpha}}_{\mu}\}$ are only functions of $g_{\mu\nu}$, hence, from the expression of $r_{\hat{0}\hat{a}\hat{b}} + r_{\hat{0}\hat{b}\hat{a}}$ (using (5-12) but in which $\{e^{\hat{\alpha}}_{\mu}\}$ is as basic variables) we see that $r_{\hat{0}\hat{a}\hat{b}} + r_{\hat{0}\hat{b}\hat{a}}$ are functions of $g_{\mu\nu}$, $g_{0i,j}$ and $g_{ij,\lambda}$. And then, according to (5-2) and (5-14), we know that $\Lambda^{\hat{a}}_{\hat{b}}$ are functions of $g_{\mu\nu}$, $g_{0i,j}$ and $g_{ij,\lambda}$ as well:

$$\Lambda^{\hat{a}}_{\hat{b}} = \Lambda^{\hat{a}}_{\hat{b}}(g_{\mu\nu}; g_{0i,j}, g_{ij,\lambda}). \tag{5-16}$$

And from (5-9) and (5-16) we know that all $\Omega^{(a)}_{\lambda}$, $e^{\lambda}_{\hat{0}}\Omega^{(a)}_{\lambda}$ and $\omega^{(a)}$ given by (5-5), (5-6) and (5-7) respectively are functions of $g_{\mu\nu}$, $g_{0i,j}$ and $g_{ij,\lambda}$ as well.

Now that the all $\Omega^{(a)}_{\lambda}$, $e^{\lambda}_{\hat{0}}\Omega^{(a)}_{\lambda}$ and $\omega^{(a)}$ are functions of $g_{\mu\nu}$, $g_{0i,j}$ and $g_{ij,\lambda}$, we can try to choose a coordinate transformation such that the metric tensor $g_{\mu\nu}$ in the new coordinate system satisfy some conditions, which can be expressed by some forms of combination of $\Omega^{(a)}_{\lambda}$, $e^{\lambda}_{\hat{0}}\Omega^{(a)}_{\lambda}$ and $\omega^{(a)}$, and we ask that these conditions lead to that there is not any second time derivative term in all the Einstein equations.

At first, for insuring that the quadratic term of time derivative in $L_{G0}$ is positive definitive, we choose (3-14), which can be generated by a special combination of $\Omega^{(a)}_{\lambda}$, $e^{\lambda}_{\hat{0}}\Omega^{(a)}_{\lambda}$ and $\omega^{(a)}$ (5-8), this means that the metric tensor $g_{\mu\nu}$ in the new coordinate system is made to satisfy (3-14).

And then, we consider the following two forms of combination of $\Omega^{(a)}_{\lambda}$:

$$W_1 \equiv 2\Omega^{(1)}_0 - \Omega^{(2)}_0 - \Omega^{(3)}_0 = 2\tilde{e}^{\hat{1}i}\tilde{e}^{\hat{1}}_{i,0} - \tilde{e}^{\hat{2}i}\tilde{e}^{\hat{2}}_{i,0} - \tilde{e}^{\hat{3}i}\tilde{e}^{\hat{3}}_{i,0} = \left(2\overline{\Lambda}^{\hat{1}}_{\hat{a}}\overline{\Lambda}^{\hat{1}}_{\hat{b}} - \overline{\Lambda}^{\hat{2}}_{\hat{a}}\overline{\Lambda}^{\hat{2}}_{\hat{b}} - \overline{\Lambda}^{\hat{3}}_{\hat{a}}\overline{\Lambda}^{\hat{3}}_{\hat{b}}\right)e^{\hat{a}i}e^{\hat{b}}_{i,0},$$

$$W_2 \equiv \Omega^{(2)}_0 - \Omega^{(3)}_0 = \tilde{e}^{\hat{2}i}\tilde{e}^{\hat{2}}_{i,0} - \tilde{e}^{\hat{3}i}\tilde{e}^{\hat{3}}_{i,0} = \left(\overline{\Lambda}^{\hat{2}}_{\hat{a}}\overline{\Lambda}^{\hat{2}}_{\hat{b}} - \overline{\Lambda}^{\hat{3}}_{\hat{a}}\overline{\Lambda}^{\hat{3}}_{\hat{b}}\right)e^{\hat{a}i}e^{\hat{b}}_{i,0};$$

From the above discussion we know that both $W_1$ and $W_2$ are functions of $g_{\mu\nu}$, $g_{0i,j}$ and $g_{ij,\lambda}$, we therefore can choose

$$W_1 = W_1(g_{\mu\nu}; g_{0i,j}, g_{ij,\lambda}) = 0, \tag{5-17}$$

$$W_2 = W_2(g_{\mu\nu}; g_{0i,j}, g_{ij,\lambda}) = 0. \tag{5-18}$$

This means that the metric tensor $g_{\mu\nu}$ in the new coordinate system is made to satisfy (5-17) and (5-18).

Under the conditions (3-14), (5-17) and (5-18), $\tilde{S}^{\hat{0}i\hat{a}}$ given by (5-15) now becomes

$$\widetilde{S}^{\hat{0}i\hat{a}} = \widetilde{e}^{\hat{a}i}\left\{\widetilde{e}^{\hat{a}k}\left[\widetilde{e}^{0}_{\hat{0}}\widetilde{e}^{\hat{a}}_{0,j} + \widetilde{e}^{j}_{\hat{0}}\left(\widetilde{e}^{\hat{a}}_{j,k} - \widetilde{e}^{\hat{a}}_{k,j}\right)\right] - \frac{1}{3}\widetilde{e}^{k}_{\hat{b}}\left[\widetilde{e}^{0}_{\hat{0}}\widetilde{e}^{\hat{b}}_{0,k} + \widetilde{e}^{j}_{\hat{0}}\left(\widetilde{e}^{\hat{b}}_{j,k} - \widetilde{e}^{\hat{b}}_{k,j}\right)\right]\right\}$$

$$= \widetilde{e}^{\hat{a}i}\left\{\frac{2}{3}\widetilde{e}^{\hat{a}k}\left[\widetilde{e}^{0}_{\hat{0}}\widetilde{e}^{\hat{a}}_{0,j} + \widetilde{e}^{j}_{\hat{0}}\left(\widetilde{e}^{\hat{a}}_{j,k} - \widetilde{e}^{\hat{a}}_{k,j}\right)\right] - \frac{1}{3}\sum_{\substack{b=1\\b\neq a}}^{3}\widetilde{e}^{\hat{b}k}\left[\widetilde{e}^{0}_{\hat{0}}\widetilde{e}^{\hat{b}}_{0,k} + \widetilde{e}^{j}_{\hat{0}}\left(\widetilde{e}^{\hat{b}}_{j,k} - \widetilde{e}^{\hat{b}}_{k,j}\right)\right]\right\},$$
(5-19)

in which there is not any time derivative term.

Based on (5-19), we can verify easily that there is not any time derivative term in (2-4) and (2-5), and there is not any second time derivative term in (2-8).

Generally speaking, a coordinate transformation can provide four coordinate conditions, but what we have used is only three of four coordinate conditions: (3-14), (5-17) and (5-18). Of course, one can try to choose other forms of combination of $\Omega^{(a)}_{\lambda}$, $e^{\lambda}_{\hat{0}}\Omega^{(a)}_{\lambda}$ and $\omega^{(a)}$.

# Appendix

$$2\left(U^k\right)_{,k} - L_V =$$

$$\frac{2\sqrt[3]{h_1}}{3\,h_1^2\,h_2^3}$$

$$(2\,h_1^2\,h_2^3\,h_{1,1,1} + 4\,h_1^2\,h_2^3\,h_4\,h_{1,1,2} + 4\,h_1^2\,h_2^3\,h_5\,h_{1,1,3} -$$
$$h_2^5\,h_{1,2,2} + 2\,h_1^2\,h_2^3\,h_4^2\,h_{1,2,2} + 2\,h_2^5\,h_3\,h_{1,2,3} +$$
$$4\,h_1^2\,h_2^3\,h_4\,h_5\,h_{1,2,3} - h_2\,h_{1,3,3} - h_2^5\,h_3^2\,h_{1,3,3} +$$
$$2\,h_1^2\,h_2^3\,h_5^2\,h_{1,3,3} + 3\,h_1\,h_2^4\,h_{2,2,2} - 6\,h_1\,h_2^4\,h_3\,h_{2,2,3} -$$
$$3\,h_1\,h_{2,3,3} + 3\,h_1\,h_2^4\,h_3^2\,h_{2,3,3} - 3\,h_1\,h_2^5\,h_{3,3,2} +$$
$$3\,h_1\,h_2^5\,h_3\,h_{3,3,3} + 3\,h_1^3\,h_2^3\,h_{4,2,1} + 3\,h_1^3\,h_2^3\,h_4\,h_{4,2,2} +$$
$$3\,h_1^3\,h_2^3\,h_5\,h_{4,2,3} + 3\,h_1^3\,h_2^3\,h_{5,3,1} + 3\,h_1^3\,h_2^3\,h_4\,h_{5,3,2} +$$
$$3\,h_1^3\,h_2^3\,h_5\,h_{5,3,3})$$

$$+\frac{1}{\sqrt[3]{h_1^2}}\,\frac{1}{18\,h_1^2\,h_2^4}$$

$$(4\,h_1^2\,h_2^4\,h_{1,1}^2 + 8\,h_1^2\,h_2^4\,h_4\,h_{1,1}\,h_{1,2} + 28\,h_2^6\,h_{1,2}^2 + 4\,h_1^2\,h_2^4\,h_4^2\,h_{1,2}^2 +$$
$$8\,h_1^2\,h_2^4\,h_5\,h_{1,1}\,h_{1,3} - 56\,h_2^6\,h_3\,h_{1,2}\,h_{1,3} + 8\,h_1^2\,h_2^4\,h_4\,h_5\,h_{1,2}\,h_{1,3} +$$
$$28\,h_2^6\,h_{1,3}^2 + 28\,h_2^6\,h_3^2\,h_{1,3}^2 + 4\,h_1^2\,h_2^4\,h_5^2\,h_{1,3}^2 + 36\,h_1^4\,h_2^2\,h_{2,1}^2 -$$
$$24\,h_1\,h_2^5\,h_{1,2}\,h_{2,2} + 24\,h_1\,h_2^5\,h_3\,h_{1,3}\,h_{2,2} + 72\,h_1^4\,h_2^2\,h_4\,h_{2,1}\,h_{2,2} +$$
$$36\,h_2^4\,h_{2,2}^2 + 36\,h_1^4\,h_2^2\,h_4^2\,h_{2,2}^2 + 24\,h_1\,h_2^5\,h_3\,h_{1,2}\,h_{2,3} +$$
$$24\,h_1\,h_2\,h_{1,3}\,h_{2,3} - 24\,h_1\,h_2^5\,h_3^2\,h_{1,3}\,h_{2,3} + 72\,h_1^4\,h_2^2\,h_5\,h_{2,1}\,h_{2,3} -$$
$$72\,h_1^2\,h_2^4\,h_3\,h_{2,2}\,h_{2,3} + 72\,h_1^4\,h_2^2\,h_4\,h_5\,h_{2,2}\,h_{2,3} + 108\,h_1^2\,h_{2,3}^2 +$$
$$36\,h_1^2\,h_2^4\,h_3^2\,h_{2,3}^2 + 36\,h_1^4\,h_2^2\,h_5^2\,h_{2,3}^2 + 9\,h_1^4\,h_2^8\,h_{3,1}^2 +$$
$$12\,h_1\,h_2^6\,h_{1,3}\,h_{3,2} - 36\,h_1^2\,h_2^5\,h_{2,3}\,h_{3,2} + 18\,h_1^4\,h_2^8\,h_4\,h_{3,1}\,h_{3,2} +$$
$$9\,h_1^4\,h_2^8\,h_4^2\,h_{3,2}^2 + 12\,h_1\,h_2^6\,h_{1,2}\,h_{3,3} - 24\,h_1\,h_2^6\,h_3\,h_{1,3}\,h_{3,3} -$$
$$108\,h_1^2\,h_2^5\,h_{2,2}\,h_{3,3} + 144\,h_1^2\,h_2^5\,h_3\,h_{2,3}\,h_{3,3} +$$
$$18\,h_1^4\,h_2^8\,h_5\,h_{3,1}\,h_{3,3} + 18\,h_1^4\,h_2^8\,h_4\,h_5\,h_{3,2}\,h_{3,3} + 36\,h_1^2\,h_2^6\,h_{3,3}^2 +$$
$$9\,h_1^4\,h_2^8\,h_5^2\,h_{3,3}^2 + 24\,h_1^3\,h_2^4\,h_{1,2}\,h_{4,1} + 60\,h_1^3\,h_2^4\,h_{1,1}\,h_{4,2} +$$
$$84\,h_1^3\,h_2^4\,h_4\,h_{1,2}\,h_{4,2} + 60\,h_1^3\,h_2^4\,h_5\,h_{1,3}\,h_{4,2} - 36\,h_1^4\,h_2^3\,h_{2,1}\,h_{4,2} -$$
$$36\,h_1^4\,h_2^3\,h_4\,h_{2,2}\,h_{4,2} - 36\,h_1^4\,h_2^3\,h_5\,h_{2,3}\,h_{4,2} +$$

$18 h_1^4 h_2^8 h_3 h_{3,1} h_{4,2} + 18 h_1^4 h_2^8 h_3 h_4 h_{3,2} h_{4,2} +$
$18 h_1^4 h_2^8 h_3 h_5 h_{3,3} h_{4,2} + 36 h_1^4 h_2^4 h_{4,2}^2 + 9 h_1^4 h_2^8 h_3^2 h_{4,2}^2 +$
$24 h_1^3 h_2^4 h_5 h_{1,2} h_{4,3} + 72 h_1^4 h_2^3 h_3 h_{2,1} h_{4,3} +$
$72 h_1^4 h_2^3 h_3 h_4 h_{2,2} h_{4,3} + 72 h_1^4 h_2^3 h_3 h_5 h_{2,3} h_{4,3} +$
$18 h_1^4 h_2^4 h_{3,1} h_{4,3} - 18 h_1^4 h_2^8 h_3^2 h_{3,1} h_{4,3} + 18 h_1^4 h_2^4 h_4 h_{3,2} h_{4,3} -$
$18 h_1^4 h_2^8 h_3^2 h_4 h_{3,2} h_{4,3} + 18 h_1^4 h_2^4 h_5 h_{3,3} h_{4,3} -$
$18 h_1^4 h_2^8 h_3^2 h_5 h_{3,3} h_{4,3} - 18 h_1^4 h_2^4 h_3 h_{4,2} h_{4,3} -$
$18 h_1^4 h_2^8 h_3^3 h_{4,2} h_{4,3} + 9 h_1^4 h_{4,3}^2 + 18 h_1^4 h_2^4 h_3^2 h_{4,3}^2 +$
$9 h_1^4 h_2^8 h_3^4 h_{4,3}^2 + 24 h_1^3 h_2^4 h_{1,3} h_{5,1} + 24 h_1^3 h_2^4 h_4 h_{1,3} h_{5,2} +$
$18 h_1^4 h_2^8 h_{3,1} h_{5,2} + 18 h_1^4 h_2^8 h_4 h_{3,2} h_{5,2} + 18 h_1^4 h_2^8 h_5 h_{3,3} h_{5,2} +$
$18 h_1^4 h_2^8 h_3 h_{4,2} h_{5,2} + 18 h_1^4 h_2^4 h_{4,3} h_{5,2} - 18 h_1^4 h_2^8 h_3^2 h_{4,3} h_{5,2} +$
$9 h_1^4 h_2^8 h_{5,2}^2 + 60 h_1^3 h_2^4 h_{1,1} h_{5,3} + 60 h_1^3 h_2^4 h_4 h_{1,2} h_{5,3} +$
$84 h_1^3 h_2^4 h_5 h_{1,3} h_{5,3} + 36 h_1^4 h_2^3 h_{2,1} h_{5,3} + 36 h_1^4 h_2^3 h_4 h_{2,2} h_{5,3} +$
$36 h_1^4 h_2^3 h_5 h_{2,3} h_{5,3} - 18 h_1^4 h_2^8 h_3 h_{3,1} h_{5,3} -$
$18 h_1^4 h_2^8 h_3 h_4 h_{3,2} h_{5,3} - 18 h_1^4 h_2^8 h_3 h_5 h_{3,3} h_{5,3} +$
$36 h_1^4 h_2^4 h_{4,2} h_{5,3} - 18 h_1^4 h_2^8 h_3^2 h_{4,2} h_{5,3} + 18 h_1^4 h_2^4 h_3 h_{4,3} h_{5,3} +$
$18 h_1^4 h_2^8 h_3^3 h_{4,3} h_{5,3} - 18 h_1^4 h_2^8 h_3 h_{5,2} h_{5,3} + 36 h_1^4 h_2^4 h_{5,3}^2 +$
$9 h_1^4 h_2^8 h_3^2 h_{5,3}^2 )$